\documentclass[journal]{IEEEtran}
\usepackage[cmex10]{amsmath}
\usepackage{array}
\usepackage{hyperref}
\usepackage{mdwmath}
\usepackage{mdwtab}
\usepackage{eqparbox}
\usepackage{url}
\usepackage{algorithm}
\usepackage{algpseudocode}
\usepackage{graphicx}
\usepackage{subfigure}
\usepackage{caption}
\usepackage{balance}
\usepackage{fancyhdr}
\usepackage{lipsum} % For filler text (optional, can be removed)

\fancypagestyle{copyright}{%
  \fancyhf{} % clear all header and footer fields
  \fancyfoot[L]{\parbox{\textwidth}{\footnotesize \textcopyright 2024 IEEE. Personal use of this material is permitted. Permission from IEEE must be obtained for all other uses, in any current or future media, including reprinting/republishing this material for advertising or promotional purposes, creating new collective works, for resale or redistribution to servers or lists, or reuse of any copyrighted component of this work in other works.}}%
}

\usepackage[a4paper, total={7in, 9in}]{geometry} % Adjust the margins to fit content better

\usepackage{color}

\begin{document}
\bstctlcite{IEEEexample:BSTcontrol}
    \title{FloRa: Flow Table Low-Rate Overflow Reconnaissance and Detection in SDN}
  \author{Ankur~Mudgal,   Abhishek Verma,~\IEEEmembership{Member,~IEEE,}
      Munesh~Singh,~\IEEEmembership{Senior Member,~IEEE,} Kshira Sagar Sahoo,~\IEEEmembership{Senior Member,~IEEE,} 
      Erik Elmroth~\IEEEmembership{ Member,~IEEE,}
      Monowar Bhuyan,~\IEEEmembership{ Member,~IEEE}%  }% <-this % stops a space
  % \thanks{Manuscript received July 10, 2012. \hl{This paper is an expanded paper from the IEEE MTT-S Int. Microwave Symposium held on June 17-22, 2012 in Montreal, Canada.} This work was funded in part by the Office of Naval Research under the Defense Advanced Research Projects Agency (DARPA) Microscale Power Conversion (MPC) Program under Grant N00014-11-1-0931, and in part by the Advanced Research Projects Agency-Energy (ARPA-E), U.S. Department of Energy, under Award Number DE-AR0000216.}
  \thanks{Ankur Mudgal and Munesh Singh are with Computer Science \& Engineering Discipline, PDPM Indian Institute of Information Technology, Design and Manufacturing Jabalpur, Madhya Pradesh, India (e-mail: 22pcso02@iiitdmj.ac.in, munesh.singh@iiitdmj.ac.in).}% <-this % stops a space
  \thanks{Abhishek Verma is with Department of Information Technology, Babasaheb Bhimrao Ambedkar University, Lucknow 226025, Uttar Pradesh, India (e-mail: abhiverma866@gmail.com).}
  \thanks{K.S. Sahoo, E. Elmorth, and M. Bhuyan are with Department of Computing Science, ADS Lab, Ume\r{a} University, Sweden. email: \{ksahoo, elmroth, monowar\}@cs.umu.se}
  }% <-this % stops a space}  
% \markboth{IEEE TRANSACTIONS ON XXXX, VOL.~60, NO.~12, DECEMBER~2012
% }{Roberg \MakeLowercase{\textit{et al.}}: High-Efficiency Diode and Transistor Rectifiers}
% ====================================================================
\maketitle
% === ABSTRACT ====================================================================
% =================================================================================

\thispagestyle{copyright}

\begin{abstract}

%---------------------------------------------------
%Software-defined networking (SDN) evolves to create next-generation networks as it enables programmability that can provision services on the fly, widely supported by OpenFlow (OF) protocol. Ternary Content Addressable Memory (TCAM) stores the flow tables in OpenFlow switches. Ternary Content Addressable Memory (TCAM) stores the flow tables in OpenFlow switches. TCAM is susceptible to various security attacks for limited storage capacity. One such attack is known as a Low-Rate Flow Table Overflow (LOFT).
%---------------------------------------------------
Software Defined Networking (SDN) has evolved to revolutionize next-generation networks, offering programmability for on-the-fly service provisioning, primarily supported by the OpenFlow (OF) protocol. The limited storage capacity of Ternary Content Addressable Memory (TCAM) for storing flow tables in OF switches introduces vulnerabilities, notably the Low-Rate Flow Table Overflow (LOFT) attacks. LOFT exploits 
the flow table's storage capacity by occupying a substantial amount of space with malicious flow, leading to a gradual degradation in the flow-forwarding performance of OF switches. To mitigate this threat, we propose \textit{FloRa}, a machine learning-based solution designed for monitoring and detecting LOFT attacks in SDN. FloRa continuously examines and determines the status of the flow table by closely examining the features of the flow table entries. Upon detecting an attack FloRa promptly activates the detection module. The module monitors flow properties, identifies malicious flows, and blacklists them, facilitating their eviction from the flow table. Incorporating novel features such as Packet
Arrival Frequency, Content Relevance Score, and Possible Spoofed IP along with Cat Boost employed as the attack detection method. The proposed method reduces CPU overhead, memory overhead, and classification latency significantly and achieves a detection accuracy of 99.49\%\, which is more than the state-of-the-art methods to the best of our knowledge. This approach not only protects the integrity of the flow tables but also guarantees the uninterrupted flow of legitimate traffic. Experimental results indicate the effectiveness of FloRa in LOFT attack detection, ensuring uninterrupted data forwarding and continuous availability of flow table resources in SDN.
\end{abstract}

 \begin{IEEEkeywords} Software Defined Networking (SDN), low-rate attack, flow table overflow.
  \end{IEEEkeywords}
%\keywords{. SDN,  low-rate attack, flow table overflow, defense system}

% \IEEEpeerreviewmaketitle

\section{Introduction}
\begin{figure}[h]
    \centering
    \includegraphics[width=.5\textwidth]{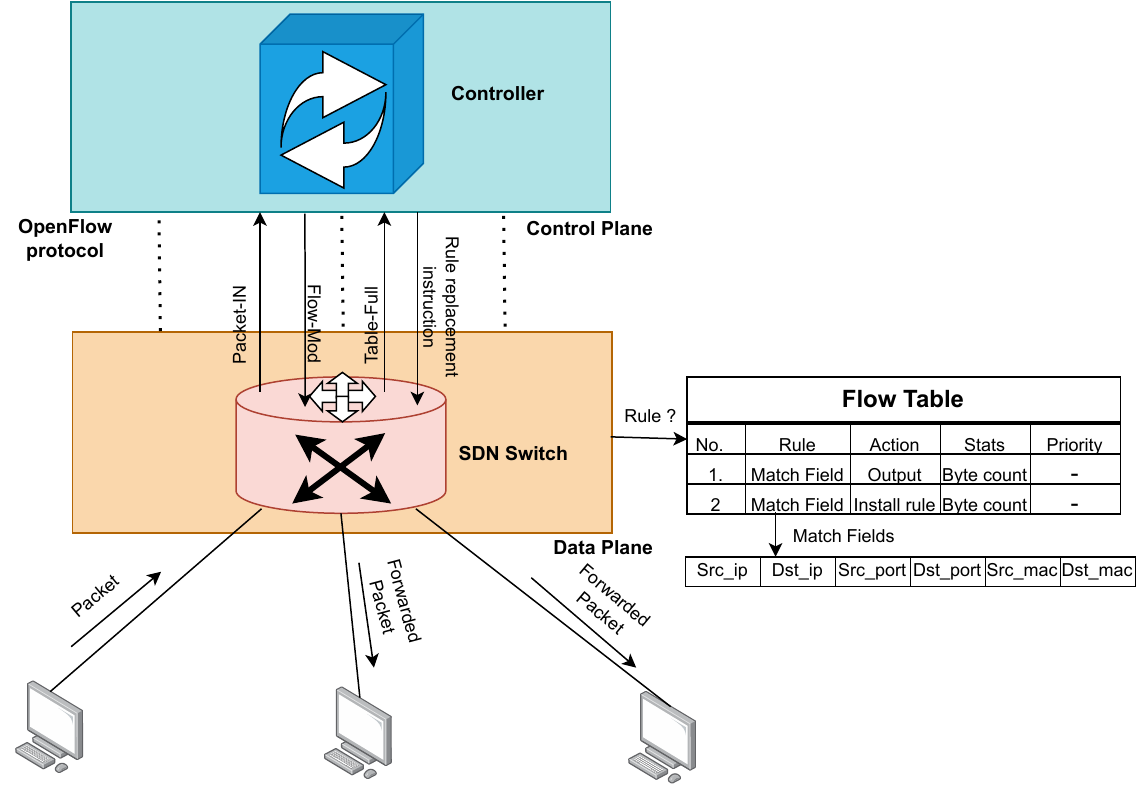}
    \caption{Overview of SDN}
    \label{SDNOverview}
\end{figure}

\IEEEPARstart{S}{DN} is a network architecture that decouples the control plane from the data plane for managing network infrastructures efficiently \cite{yan2015software}. Instead of requiring manual configuration for each network device, SDN enables network administrators to manage and administer the network using a programmatic approach. A high-level SDN design is depicted in Fig. \ref{SDNOverview}. Although SDN enhances the performance of traditional networks, there exist critical security flaws inherent in its design. It is particularly susceptible to flow table overflow in the SDN data plane (or SDN switch). Since switches in SDN lack intelligence, if a flow does not comply with any of the entries in the flow table, a packet-in request is generated to trigger a new flow rule from a logically centralized controller. However, an attacker may exploit this approach to install new rules in the switch by sending specially crafted packets that consume the flow table buffer. To achieve high access performance, the recent SDN switches  handle less number of flow entries, e.g., a few thousand rules \cite{katta2016cacheflow}, \cite{cao2022attack}, \cite{lu2011serverswitch}. \textcolor{black}{Additionally, PICA8 has published a white paper highlighting the capacity limitations of Trident chipset-based SDN switches, typically limited to a few thousand rules. Moreover, the discussed attack possesses a stealthy and undetectable nature, enabling it to overflow switches of larger sizes as well. Therefore, increasing switch capacities is not a viable long-term, cost-effective or permanent solution for mitigating such attacks.} Ternary Content Addressable Memory (TCAM) stores flow rules in SDN switches. Due to its high cost and power-hungry nature, TCAM of limited capacity is employed in real-world network deployment \cite{rawat2016software}.

The attacks demonstrated in the literature often produce a significant number of small, periodic flows every second to overload the SDN switches successfully \cite{tang2023ftop}, \cite{tang2023ltrft}, \cite{qian2016openflow}. 
\textcolor{black}{The existing solutions account for a limited number of attack packets. However, the substantial overall attack packet rate is capable of overwhelming the flow table \cite{qian2016openflow}, \cite{shang2017flooddefender}, \cite{wang2015floodguard}.}
%These attacks are easy to detect and can be addressed by Available solutions considers a limited number of attack packets. Still, the overall attack packet rate is significant to overload the flowtable 
% \mb{*incomplete sentence*} \cite{qian2016openflow}, \cite{shang2017flooddefender}, \cite{wang2015floodguard}. 
Although existing works attempt to lower attack rate, unfortunately, they failed in actual (practical) SDN settings because they did not consider the flow table's specific details, such as the flow rules duration \cite{pascoal2017slow}. Moreover, some attacks call for the compromise of a huge number of hosts, which comes at a high cost to the attacker \cite{pascoal2017slow}. However, specially designed low-rate attacks are potential threats to SDN. The flow tables can easily be consumed due to their limited storage ability, which makes the switches unable to store new legitimate rules, thus causing performance degradation of SDN. Besides, the new incoming packets are forwarded to the controller, which, as a result, increases the risk of data-control saturation attacks \cite{swami2019software}.
%%%%%%%%%%%%%%%%%%
%% mb --- finished
%%%%%%%%%%%%%%%%%%

Some attacks target the storage ability of flow tables \cite{cao2018disrupting}, \cite{tang2023ftmaster}, \cite{cao2022attack}, \cite{yu2021flow}. The attacker sends the forged packets in fixed intervals regularly to install new rules and consume the flow table storage space. Because of the lower attack rate, that attacker hides itself within the normal flow, which makes it stealthy and undetectable. The prevalent countermeasures against LOFT attack \cite{cao2022attack} at the moment are not adequately explained and mainly concentrate on dealing with flow table overflow as a whole without going into the specifics of flow table entries, like rate restriction \cite{xu2017mitigating}, flow scheduling \cite{pei2018resource}, and replacement policy betterment \cite{pascoal2020slow}. Along with the advancement of attack detection, flow intrinsic features have been proposed by various researchers from a medium-grained perspective, but they are not extensive \cite{xie2018line}, \cite{liang2019industrial}. Similarly, some previous studies focused on increasing flow table use without taking malicious attack situations into account \cite{shirali2018delayed}. Most existing work in the literature focuses on high-rate attacks, and very few studies have been done to analyze low-rate attacks with similar impact on SDN security, which is one of the important research gaps in this field. 

This paper addresses a LOFT attack that can composedly consume the flow table storage with a low rate of attack packets. To launch such attacks, the attacker needs to design the attack packets in such a way that it can install new flow rules and does not let the controller delete the existing attack rules that eventually overflow the flow table and cause performance degradation on the SDN switch while maintaining a low packet sending rate.
% \textcolor{black}{To achieve this, the attacker needs to have the understanding of rule installation strategy. As SDN controller is responsible for designing the rule installation logic, which is not directly accessible by the attacker. Also newly installed rules will be evicted after the stipulated time due to timeout setting. So the attacker need to have understanding of timeout setting as well.}
 Given the above, FloRa is proposed as a detection method to address LOFT attacks against SDN. %FloRa consist of four modules which perform flow rule analysis, anomaly prediction, feature extraction and detection tasks. Flow analyzer polls the flow table within a certain duration from it's current state and activates Anomaly predictor upon detection of exceeding flow rules beyond user defined attack detection threshold. The Anomaly predictor then analyzes the data and activates Feature extractor module, that extracts the important features. Finally the detection module performs classification to predict the attack using Machine Learning (ML). To evaluate the performance of FloRa, experiment are carried out and the results shows that the algorithm performs well in detection of LOFT attacks.
 The summarized contributions are as follows: 
\begin{itemize}
    \item This paper suggests FloRa, a modular and scalable SDN data-plane security framework, to detect and mitigate LOFT attacks using machine learning techniques.
    \item FloRa consist of four modules: Flow analyzer, Anomaly predictor, Feature selector and Attack detection.
    \item In this paper, three novel features, PAF, PSI and CRS, are introduced for the first time with regular statistical features.
    \item In this paper, BCP38 is used for analyzing spoofed IP, RFECV is used to calculate feature importance and Cat Boost is employed for detection.
\end{itemize}

The structure of the paper is as follows. Section \ref{bg} provides information about the background of the research area. Section \ref{rw} reviews the related works in the literature. Section \ref{ap} describes the important prerequisites for launching a low-rate Distributed Denial of Service (DDoS) attack against SDN. The Threat model is discussed in Section \ref{tm1}. In Section \ref{pm}, a detailed discussion on the functionality of FloRa is carried out. Section \ref{ee} describes the experimental methodology and performance evaluation. Lastly, the paper is concluded in Section \ref{cc}. 

Table \ref{tab:rel} contains the complete form of the abbreviation symbols used in this paper.

\begin{table}[ht]
\renewcommand{\arraystretch}{1.5} % Increase the vertical spacing by 50%
\centering
\caption{Symbol Table}
\begin{tabular}{|c|c|}
 \hline
 Abbreviation & Full Form \\
 \hline
  LOFT & Low rate flow table overflow\\
 \hline
 TCAM  & Ternary content addressable memory\\
 \hline
 OF & OpenFlow\\
 \hline
 AF & Attack frequency\\
 \hline
 ANP  &  Added new packets\\
 \hline
 RPR  &  Regulated packet rate\\
 \hline
 MRI  &  Mean rate of increase\\
 \hline
 PAF  &  Packet arrival frequency\\
 \hline
 PSI  &  Possibility of spoofed IP\\
 \hline
 CRS  &  Content relevance score\\
 \hline
 RTT  &  Round trip time\\
 \hline
 IG  &  Information gain\\
 \hline
 H  &  Entropy\\
 \hline
 
\end{tabular}
    \label{tab:rel}
\end{table}

\section{Background}\label{bg}
\subsection{SDN}
SDN decouples the data plane and control plane to enable centralized control. While the data plane manages the forwarding mechanism of network packets, the forwarding logic is to be designed by the centralized controller. Programmability is another important feature of SDN. Software-based controllers make it easier for the network administrator to configure the required changes in the network. SDN enables vendor independence, as the hardware and software are separated. Most importantly, security measures can be implemented efficiently in SDN due to its ability to have a global view of the network. To achieve all the above-discussed functionalities, communication between control and data plane need to be established, which is done by OF protocol \cite{tourrilhes2014sdn}. 

\subsection{OpenFlow}

OpenFlow (OF) is a network protocol and open standard, which is a vital component of SDN. OF protocol \cite{tourrilhes2014sdn} is the first and widely adopted protocol that enables communication between controller and network switches using Southbound API. OF allows network devices to process packets based on the content defined in the flow table, which the controller generates and manages. A flow table comprises flow entries, known as rules, consisting of match fields and respective actions, along with other details. When a packet arrives at the SDN switch, its header is compared with the existing flow rules stored in the flow table. If there is a match, a corresponding action is taken, like sending further or deleting the packets.

The functionality of SDN is illustrated in Fig. \ref{SDNOverview}. When a packet arrives at the SDN switch, it is matched with the flow table entries to determine whether a rule exists for the incoming packet. If no match is found, the packet is treated as a new packet, triggering the Packet-In message to notify the controller, which can take the appropriate action. Subsequently, the controller instructs the switch using Flow-Mod to install a new flow rule for the incoming packet. The switch then checks the capacity of the flow table; if the table is full, it sends a packet to the controller informing that it is full. This signal prompts the controller to remove rules following predefined logic, freeing up memory. At last, the new rule will be installed.
\subsection{\textcolor{black}{Low-rate DDoS}}
\textcolor{black}{Low-rate DDoS attack, present a sophisticated approach to distributed denial-of-service (DDoS) strategies, with the primary goal of gradually and covertly overwhelming network resources. Unlike high-volume DDoS attacks, which inundate the target with a sudden influx of traffic, Low-rate DDoS attacks operate by gradually exhausting network resources over an extended period. Attackers leverage low-rate traffic to target vulnerabilities in network protocols or applications, gradually exploiting weaknesses such as bandwidth, CPU, and memory. This gradual approach makes detection challenging for conventional mechanisms, as the traffic volume remains below typical threshold levels. In contrast, high-volume DDoS attacks deliver a massive volume of traffic in a short timeframe, causing immediate disruption by overwhelming network resources. Although high-volume attacks are more conspicuous and can be detected and mitigated relatively quickly, Low-rate DDoS attacks present a significant challenge due to their subtle and prolonged nature. Moreover, the prolonged duration of Low-rate DDoS attacks increases the likelihood of remaining undetected, allowing attackers to sustain the assault and inflict prolonged damage. As a result, combating Low-rate DDoS attacks necessitates advanced detection techniques capable of identifying subtle changes in network behavior and distinguishing malicious traffic from legitimate patterns.}

\subsection{Attack Parameters}
To carry out the LOFT attack, some parameters need to be taken care of by the attacker. Such crucial parameters are discussed below.

\subsubsection{Attack Frequency (AF)}
To implement the attack, it is required to ensure that the switch does not delete the installed attack flow rules due to the timeout setting of the OF switch. To achieve this, the attacker retransmits the attack packets within a specific interval which is known as AF. For example, considering the idle timeout of 20 seconds, the attack packets should be typically retransmitted between 5 and 19 seconds.

\subsubsection{Added New Packets (ANP)}
The basic idea behind the LOFT attack is to consume the capacity of the flow table. The attacker gradually increases the attack packets and the initially sent packets to increase the attack flow rules. The increased number of attack packets among adjacent AF is called ANP. The ANP should be set so that the attack remains stealthy.

\subsubsection{Regulated Packet Rate (RPR)}
To make the attack effective, the attack must remain undetectable. The transmission rate of attack packets is set so that the flow table's storage capacity can be consumed, which leads to overflow. For example, assume the switch has $p$ ports, and flow arrival frequency is $q$ per second. So, the flow table's used capacity can be calculated using Eq. \ref{cap}.

\begin{equation}
  C_{\text{used}} = p \cdot q \cdot T_{\text{idle}}
  \label{cap}
\end{equation}
where, $C_{\text{used}}$ represents used capacity, $p$ is the number of ports, $q$ represents packet arrival frequency, and $T_{\text{idle}}$ is the  idle timeout value.

\noindent With the help of ANP and AF, the mean rate of increase (MRI) in the attack rule installation can be calculated as Eq. \ref{mri}.

 \begin{equation}
     MRI = \frac{ANP}{AF}
     \label{mri}
 \end{equation}
% where, $MRI$ represents mean rate of increase, $ANP$ indicates added new packets, and $AF$ represents arrival frequency of packets.
 
\noindent Now, the total duration of attack required to overflow the table can be calculated as Eq. \ref{d}.

 \begin{equation}
     D_{\text{total}} = \frac{C - C_{\text{used}}}{MRI}
     \label{d}
 \end{equation}
where, $D_{\text{total}}$ is the total duration of attack, $C$ resembles capacity of flow table, $C_{\text{used}}$ indicates the used capacity of flow table, and $MRI$ represents the mean rate of increase. 

To ensure the effectiveness of the attack, all the parameters should be set so that the standard setting, like idle timeout, does not delete the installed attack flow rules. RPR must also be higher than the flow table's capacity limit so that overflow can still be pursued even if some rules are eliminated.

\section{Related Work}\label{rw}
OF switches make up the SDN data plane, making it susceptible to different security attacks. DDoS attackers can deplete the resources of controllers and switches by specially designed packets sent intentionally at a low rate. Currently, there are two primary kinds of attack detection solutions for SDN networks against low-rate attacks (i.e., feature-based and threshold-based), which are reviewed in this section. 

\subsection{Feature based detection}
The main goal of this method is to develop a classifier that can efficiently differentiate normal and attack streams. Typically, attack signatures are processed, and subsequent detection models are built using ML and statistical analysis. 

Tang et al. \cite{tang2023ftmaster} suggested a solution for the LOFT attack in SDN. Firstly, the network features under low-rate attacks are analyzed, and then the features that are found effective in identifying low-rate attacks are selected. XGBoost is used as a detection classifier, and the selected features are fed to the model for training. Experiments demonstrate that the method has an acceptable detection rate. However, this method imposes overhead in terms of classification time and memory usage that can delay detection, causing the attack's severity. A method that addresses LOFT attacks over SDN is proposed based on the HGB-FP method by Tang et al. \cite{tang2021real}. The method wisely integrates fuzzy logic with ML framework using HGB-FP (Honeybee Genetic-Based Fuzzy Petrinet) for attack detection; it offers an adaptive response that allows the method to adjust the mitigation strategies dynamically based on threat levels. Multiple sets of experiments are carried out to verify the performance of the algorithm. In addition, the algorithm is compared with other detection algorithms and found to be performing well. However, the method suffers from some limitations. The method involves an adaptive response mechanism, making it resource-intensive. Also, the method employs HGB-FP which requires specialized expertise and resources for implementation and maintenance. 

Tang et al. \cite{tang2023ftop} proposed a method for detecting low-rate DDoS attacks on SDN framework that employs Random Forest for detection and mitigation of the attack and achieves 98.89\%\ accuracy. FTOP adopts a proactive approach by implementing preventive measures to avoid flow table overflows before they occur. The system employs adaptive algorithms to allocate resources and manage flow entries dynamically, optimizing the usage of limited resources. However, training a Random Forest model can be costly in terms of computation, especially when working with many trees or features. The prediction stage may also be somewhat sluggish in real-time applications. Despite their robustness, Random Forests still need hyperparameters like the number of trees, maximum depth, and minimum leaf samples to be carefully tuned to function at their best. Correlated features could be rugged for Random Forests to handle. In the case of multicollinearity, the relevance scores they offer could not accurately reflect the true value of specific variables. In work by Perez-Diaz et al. \cite{perez2020flexible}, six different ML techniques are used to classify low-rate attacks. Given that the word "flexible" connotes flexibility and adaptability, the architecture was planned to support various ML models to address LOFT attacks. The modules that need high computation are kept outside the network. The proposed intrusion detection system (IDS) communicates with the controller using platform-independent identification APIs. There are certain limitations associated with the proposed solution. The packet is forwarded to IDS, resulting in high transmission and propagation delay. It is crucial to choose the right ML models. For various elements of the problem, the performance of different models, such as decision trees (DT), random forests (RF), support vector machines, or deep learning neural networks, may vary. Using multiple ML techniques also results in higher detection time and consumes the CPU. Muraleedharan et al. \cite{muraleedharan2021deep} suggested using RF, XGBoost, and DT models for evaluation. The authors tested their models against two benchmark datasets for security, and the results showed that the suggested classifiers had an accuracy of more than 99\%. However, the paper suffers from some limitations. Firstly, the proposed solution specifically targets HTTP get and post-attack. Secondly, dataset-oriented testing does not analyze the system's response time and scalability of the approach. Also, using three different models, which are resource-intensive, may lead to classification delay. Zhijun et al. \cite{zhijun2020low} proposed a solution for low-rate attacks using a 'Factorization Machine', an ML algorithm. The method extracts four features from the dataset and applies the algorithm to it for attack detection. However, the accuracy, precision and recall achieved by the method can be improved further. Also, the number of features employed for the detection dataset is less, and there exists a chance of misclassification. 

\subsection{Threshold-based detection}

In addition to the above-mentioned works, some works focused on development of the solutions that implement the statistical use of dynamic thresholds for differentiating between normal and malicious traffic patterns in SDN when low-rate DDoS attack encounters. 

K. Hong et al. \cite{hong2017sdn} proposed an SDN-assisted detection method for low-rate attacks. The paper deals with addressing HTTP POST attacks. In this paper, the authors have provided an approach that resolves slowloris and HTTP POST attacks on SDN with the help of the Slow HTTP DDoS Defense Application (SHDA) that runs on the SDN controller. Based on timeout, the SHDA processes the packets and determines if the traffic is malicious. However, the paper suffers from some serious limitations. Firstly, the low-rate attack on OF switch TCAM is not addressed here in the paper. Secondly, the method for calculating the threshold is not explained clearly. Zhaou et al. \cite{zhou2018exploiting} propose a solution for addressing the LOFT attack. The proposed method addresses the low-rate attack on OF switch TCAM. The TCAM capacity and timeout inference mechanism is implemented to address the attack. The inference attack is implemented by recording the RTT of the packets. This paper also suffers from limitations. Firstly, the TCAM memory is very expensive and increasing the TCAM capacity will be an additional overhead. Also, the proposed solution is not feasible for all platforms. Dhawan et al. \cite{dhawan2015sphinx} proposed a method SPHINX that offers in-the-moment network behaviour verification. It actively monitors all pertinent OF control messages from the switches and the controller. SPHINX can maintain continuous visibility into network activity due to real-time monitoring capability. SPHINX utilizes a custom algorithm designed explicitly for fast validation of network updates as the controller processes. This custom algorithm is likely tailored to the unique requirements and characteristics of SDN environments, allowing for rapid and accurate verification of network changes. However, the approach suffers from some limitations, and it can be challenging to maintain the system's efficiency in large-scale SDN implementations. It is crucial to ensure it can manage a sizable number of switches, controllers, and network traffic. Even with real-time monitoring, detection lag may exist, mainly when dealing with complicated network dynamics. Timely responses to security concerns depend on reducing latency. Cao et al. \cite{cao2022attack} proposed LOFTGaurd to effectively detect LOFT attacks, which target SDN flow tables by consuming their storage capacity with a low rate of malicious traffic. It seeks to keep SDN switches operating correctly even under attacks. LOFTGuard data plane components implemented in C as daemons operate alongside the original OF agent on the switch that helps analyse the flow. With this implementation method, there is no need to change the hardware or recompile the OF agent. However, the performance of the method is limited to some extent. LOFTGuard software is designed to operate with the EdgeCore AS4610-54T switch's particular hardware. The hardware capabilities of this switch may have an impact on its efficacy. To move flow rules across various places, dependent on how live they are, the system uses a set liveness threshold of 200 ms. This threshold might not adjust to changes in the network in real-time.

\textcolor{black}{Understanding the evolving nature of these attacks and developing an effective solution poses a significant challenge. Identifying the origins of such attacks is complicated due to their prolonged duration and diverse sources. To manage the flow table effectively and secure the controller, we create test datasets that characterize the LOFT attack.
%Addressing this necessitates improved data flow management in SDN switches, bolstering the security of SDN controllers, and generating comprehensive test datasets illustrating the characteristics of LOFT attacks. 
Recognizing the need for further research and attention, we propose FloRA as a solution. This approach addresses various limitations, including classification delay, false negative rates, memory overhead, and heavy CPU usage, offering an innovative solution for detecting low-rate DDoS attacks in SDN.}

 %Figuring out how these attacks change over time and coming up with ways to stop them as they change is a big challenge. It's also hard to figure out where these attacks are coming from because they last a long time and come from many different sources.  Hence proposing the solution FloRA that considers the limitations like classification delay, false negative rate, memory overhead, heavy CPU usage and provide a novel solution for detection of low rate DDoS attack in SDNs.

\section{Attack Preparation}\label{ap}
To overflow the flow table of the OF switch, the attacker must craft packets so that a different flow rule can be installed for each new packet. Simultaneously, the newly installed rule should be retained in the flow table for an extended duration, making it challenging for legitimate hosts to get the service, i.e., causing a DDoS attack. To accomplish this, the attacker needs to assess the configuration settings of the installed OF switch.

\subsubsection{Analysis of match fields}
Standard match fields are associated with every incoming packet in an OF switch. Modifying these fields allows us to assess the potential for triggering the creation of a new flow rule in the flow table. Algorithm 1 outlines the steps involved in analyzing these match fields. The primary objective of Algorithm 1 is to examine the fields ($f$) within network packets that have the potential to initiate the creation of a new flow rule during the matching process. The algorithm commences by initializing an empty list ($F$) to collect identified fields. It then proceeds to construct a new packet and transmit it to a specific destination. After receiving a response from the destination, we record the 'Round Trip Time (RTT)' as $T_0$. Subsequently, the same packet is resent to the same destination, and the RTT for this transmission is recorded as $T_1$. To gain insights, we compare the values of $T_0$ and $T_1$. A larger value for $T_0$ compared to $T_1$ signifies that during the initial packet transmission when there is no entry in the table for the corresponding packet, the packet's information is sent to the controller for appropriate action. However, during the second transmission, the packet is directly routed to the destination due to the existence of an established flow rule entry.

To further analyze, a new packet is crafted by modifying the specific field ($f$) and sent to the same destination. Its RTT is recorded as $T_2$. Next, we compare the values of $T_0$ and $T_2$. If $T_0$ is found to be equal to $T_2$, it indicates that the altered field ($f$) triggered a new flow rule, and this specific field ($f$) will be added to the list. The above-discussed process will be repeated $N$ times to validate the fields and analyze other fields that can trigger the new flow rules. 

\begin{algorithm}[!h]
\caption{Pseudocode of Match Fields Analysis}\label{alg:analysis}
\begin{algorithmic}[1]
\Procedure{AnalysisOfMatchFields}{} 
\State $F \gets [\,]$
\State $f \gets matchfields$
\State Craft a packet and send it
\State $pkt_0 \gets$ craft pkt($\text{dest}$, $f$)
\State $F.\text{append}($Transmit pkt($pkt_0$).$f$$)$
\State Record RTT as $T_0$
\State Resend the packet and record RTT as $T_1$
\If{$T_0 > T_1$}
    \State Initialize an empty list $M \gets [\,]$
    %\State $f \gets 0$
    %\While{$f < \text{length}(F)$}
        \State $pkt_1 \gets$ craft pkt($\text{dest}$, $F[f]$)
        \State Initialize empty lists $T_2 \gets [\,]$
        \State $i \gets 0$
        \While{$i < \text{length}(F)$}
            \State Transmit pkt($pkt_1$)
            \State $pkt_i \gets$ modify\_matchfield($pkt_1$, $F[i]$)
            \State $F.\text{append}($Transmit pkt($pkt_i$)$)$
            \State $T_2.\text{append}($Transmit pkt($pkt_i$).RTT$)$
            \If{$T_{2} \approx T_0$}
            \State $M.\text{append}(F[f])$
        \EndIf
            \State $i \gets i + 1$
        \EndWhile
        % \State $\overline{T}_2 \gets$ average($T_2$)
        %  \State $\overline{T}_3 \gets$ average($T_3$)
        % \If{$\overline{T_{2}} \approx T_0$}
        %     \State $M.\text{append}(F[f])$
        % \EndIf
        %\State $f \gets f + 1$
    % \EndWhile
\EndIf
\EndProcedure
\end{algorithmic}
\end{algorithm}

\subsubsection{Analysis of Idle Timeout}
There is a timeout setting configured in the OF switch, which evicts installed flow rules from the flow table after a specific duration of inactivity, known as the idle timeout. The attacker needs this information so that packets can be resent within the idle timeout duration to refresh it and keep the flow rules retained for a more extended period. The steps to analyze the idle timeout are outlined in Algorithm 2. The algorithm begins by creating and sending a packet denoted as $pkt$ to calculate the RTT as $t_0$. This packet includes a specific destination address $dst$ and the match fields represented as $M$, which were inferred in Algorithm 1. The same packet is resent again, and RTT is recorded as $t_1$. Now, the recorded RTTs are compared, and an empty list $t$ is created. The same packet will be sent with an increasing interval of 1 second, and RTT will be recorded as $t_i$. The same packet will be sent repeatedly till the value of $t_i$ differs from $t_1$.The recorded RTTs are compared with $t_0$ every iteration. If a value of $t_i$ is equal to $t_0$, it will be added to $t$. The experiment is repeated $n$ times for better accuracy. ANOVA is employed for statistical analysis of the recorded RTTs to understand their differences.

%\textbf{Total Variation (Total Sum of Squares, SST):}
%\begin{equation*}
%SST = \sum\sum (RTT_{ij} - \text{Grand Mean})^2
%\end{equation*}

%\textbf{Between-Group Variation (Between-Group Sum of Squares, SSB):}
%\begin{equation*}
%SSB = \sum (\sum (RTT_{ij} - \text{Group Mean})^2)
%\end{equation*}

%\textbf{Within-Group Variation (Within-Group Sum of Squares, SSW):}
%\begin{equation*}
%SSW = \sum\sum (RTT_{ij} - \text{Group Mean})^2
%\end{equation*}

%\textbf{Degrees of Freedom (DF):}
%\begin{align*}
%DF_{\text{Total}} &= N - 1\\ 
%DF_{\text{Between}} &= k - 1\\ 
%DF_{\text{Within}} &= N - k\\
%\end{align*}

%\textbf{Mean Square (MS):}
%\begin{align*}
%MS_{\text{Between}} &= \frac{SSB}{DF_{\text{Between}}} \\
%MS_{\text{Within}} &= \frac{SSW}{DF_{\text{Within}}}
%\end{align*}

%\textbf{F-Statistic (Test Statistic):}
%\begin{equation*}
%F = \frac{MS_{\text{Between}}}{MS_{\text{Within}}}
%\end{equation*}

%\textbf{p-Value:}
%\begin{equation*}
%\text{p-value} = \text{F-distribution with } DF_{\text{Between}} %\text{ and } DF_{\text{Within}}
%\end{equation*}%

The resultant value of the ANOVA test is taken as the idle timeout.

\begin{algorithm}[!h]
\caption{Pseudocode of Idle Timeout Estimation}
\begin{algorithmic}[1]
% \REQUIRE $dst$: Target destination, $pkt$: Initial packet, $max\_int$: Max resend interval, $\alpha$: Significance level, $max\_iter$: Max iterations
% \ENSURE $t_\text{idle}$: Estimated idle timeout

\State $t_0 \leftarrow$ record\_RTT($pkt$, $dst$, $M$)
\State $t_1 \leftarrow$ resend\_and\_record\_RTT($pkt$, $dst$, $M$)

\If{$t_0 > t_1$}
    \State $t \leftarrow [t_1]$
    \State $interval \leftarrow 1$
\Else
    \State $t \leftarrow [\,]$
    \State $interval \leftarrow$ None
\EndIf

\State $iter \leftarrow 0$
\While{$iter < max\_iter$}
    \If{$interval \neq$ None}
        \State $t_i \leftarrow$ resend\_and\_record\_RTT($pkt$, $dst$, $M$)
        \If{$t_i \approx t_0$}
            \State $t$.append($t_i$)

            \State $p \leftarrow$ ANOVA\_test($t$)

            \If{$p \leq \alpha$}
                \State $t_\text{idle} \leftarrow t_i$
                \State \textbf{break}
            \EndIf
        \EndIf
    \EndIf

    \State $interval \leftarrow interval + 1$
    
    \If{$interval > max\_int$}
        \State \textbf{break}
    \EndIf
    
    \State $iter \leftarrow iter + 1$
\EndWhile

\If{$t_\text{idle}$ is not assigned}
    \State $t_\text{idle} \leftarrow$ None
\EndIf

\Return $t_\text{idle}$
\end{algorithmic}
\end{algorithm}

\section{Threat Model} \label{tm1}
\begin{figure}
    \centering
    \includegraphics[width=0.35\textwidth]{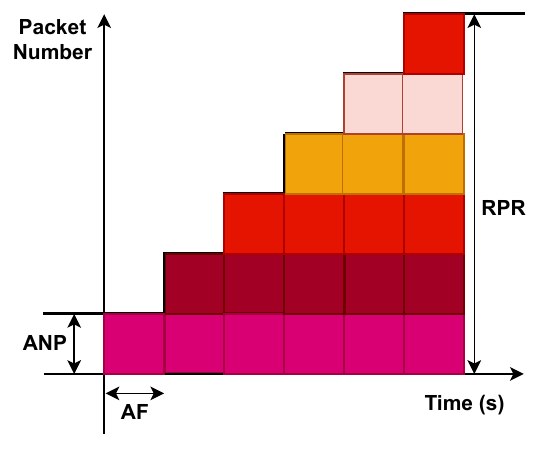}
    \caption{Threat Model}
    \label{tm}
\end{figure}

The threat model for LOFT attacks, illustrated in Figure \ref{tm}, encompasses three critical parameters: Attack Frequency (AF), Added New Packets (ANP), and Regulated Packet Rate (RPR). When a packet is retransmitted, the AF determines the time delay, which serves as the signal for flow entry matching. The AF must be less time than the idle timeout to avoid timeout mechanisms. As the number of packets transmitted between nearby AFs increases, AF is essential for gradually overloading the flow table. Due to ANP, more attack flow entries are produced when there are more AFs. For a successful overflow, the attacker must submit more attack flow entries than the targeted flow table's remaining capacity, represented by the RPR. LOFT attacks maintain a shallow attack rate in contrast to conventional high-rate attacks. Unmatched packets introduced by these attacks cause flow entry installations, damaging SDN switches. Attackers resend duplicates throughout each idle timeout period to avoid eviction due to timeouts, a tactic that can easily blend in with background traffic. This method eventually causes flow tables to overflow, which makes it challenging to provide consistent services to legitimate flow entries. 

\section{Proposed Method}\label{pm}
\subsection{System Overview}
\begin{figure}[ht]
    \centering
    \includegraphics[width=.5\textwidth]{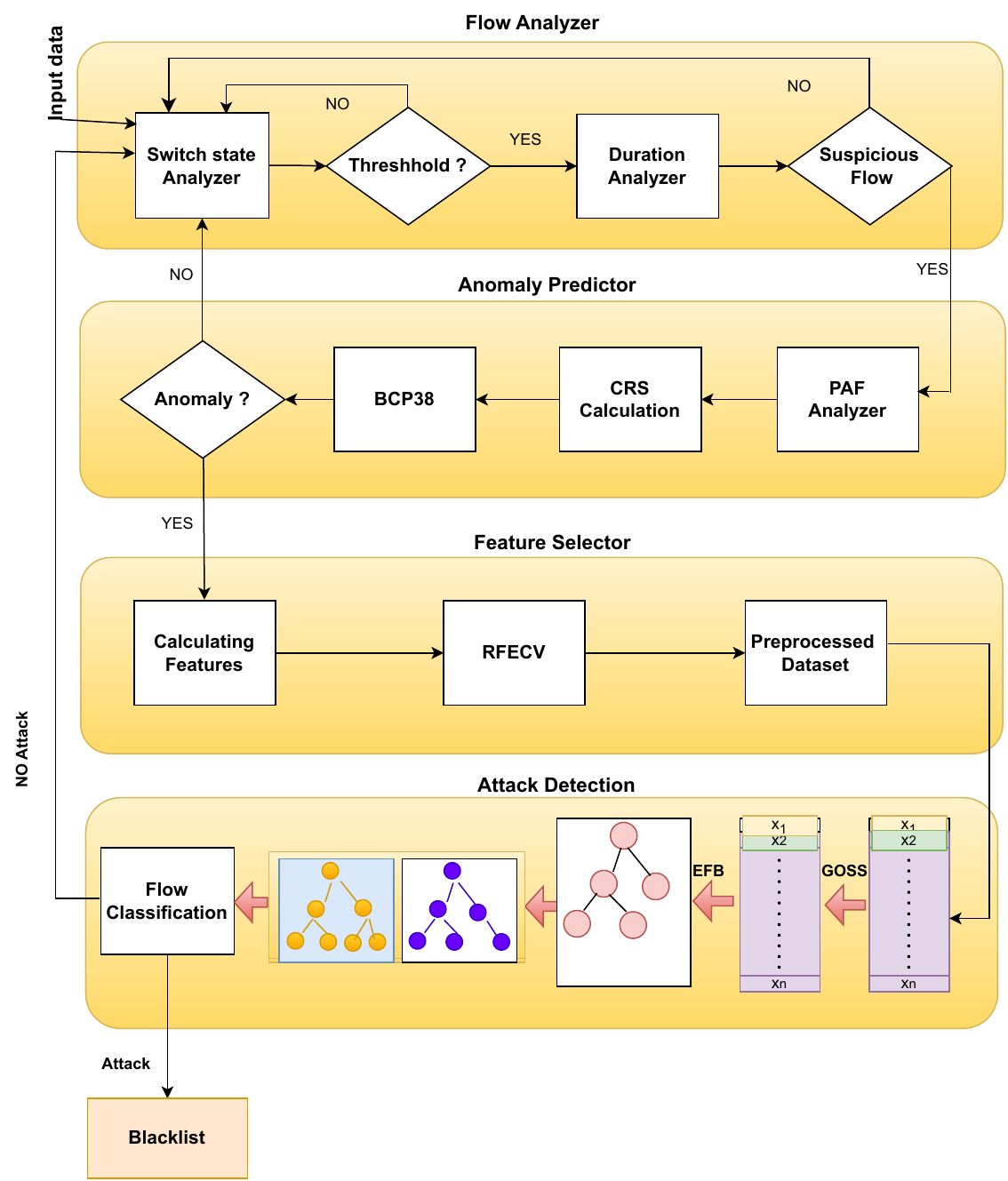}
    \caption{Framework of FloRa}
    \label{FloRa}
\end{figure}

FloRa comprises four modules: the flow analyzer, anomaly predictor, feature selector, and attack detection. The proposed framework is illustrated in Fig. \ref{FloRa}. The flow analyzer module is responsible for several actions, including monitoring the most recent state of the flow table, counting flow entries in the table, and identifying suspicious flows. If a suspicious flow is detected, the flow analyzer module activates the anomaly predictor module. The anomaly predictor module is responsible for analyzing the pattern of packet arrivals, checking the validity of the packets, and assessing the possibility of spoofed IP addresses. The anomaly predictor module invokes the feature selector module upon detecting any anomalies. The feature selector module computes features from the provided data and then calculates feature importance. The selected features are then provided as input to the detection module. This module categorises the flow into attacks, and benign as well as identified attack flows are added to the blacklist. 

\subsection{Flow Analyzer}
The flow analyzer module first counts the number of flow rules present in the flow table at the current state. If the number of flow rules present exceeds the specified threshold value, the switch state analyzer calls the duration analyzer module. The duration analyzer checks the duration for which the rules reside in the flow table. Any rules residing for more than 100 seconds are marked as suspicious flows. As per \cite{zhijun2020low}, the average lifetime of a legitimate flow rule is $10$ seconds. In a typical data center, only $0.1\%$ of flow rules are retained for $200$ seconds, while $80\%$ of flow rules are evicted after the idle timeout duration. Based on this study, a duration threshold of $100$ seconds is considered to avoid false positives. The switch state analyzer is responsible for maintaining the most recent data on the flow rules. It takes samples regularly, typically aligned with the idle timeout duration. A threshold is used in this module to predict potential attack risks. The choice of the threshold value should strike a balance; it should not be set too high, as this might delay the detection of an attack, nor too low, as this might lead to unwanted overhead.

\subsection{Anomaly Predictor}
The anomaly predictor analyzes the flow rule data and predicts possible anomalies by calculating packet arrival frequency, content relevance score, and possibility of the spoofed IP.

\subsubsection{Packet Arrival Frequency (PAF)}
The packet arrival frequency is calculated by analyzing the duration for which the flow rule is present in the flow table and the number of packets transmitted in this duration. Eq. \ref{lambda} describes the calculation used for packet arrival frequency. 

\begin{equation}
\lambda = \frac{\sum_{i=1}^{n} \text{AT}_{i} - \text{AT}_{j}}{N}
\label{lambda}
\end{equation}
\textcolor{black}{where, $\lambda$ represents packet arrival frequency, $n$ represents number of observation interval, $AT_i$ represents packet arrived in the $i^{\text{th}}$ interval and $AT_j$ denotes the packet arrived at $j^{\text{th}}$ interval, $N$ denotes total number of observation.}

\begin{figure}[ht]
    \centering
    \subfigure[Probability distribution for normal traffic\label{normal}]{\includegraphics[width=.4\textwidth]{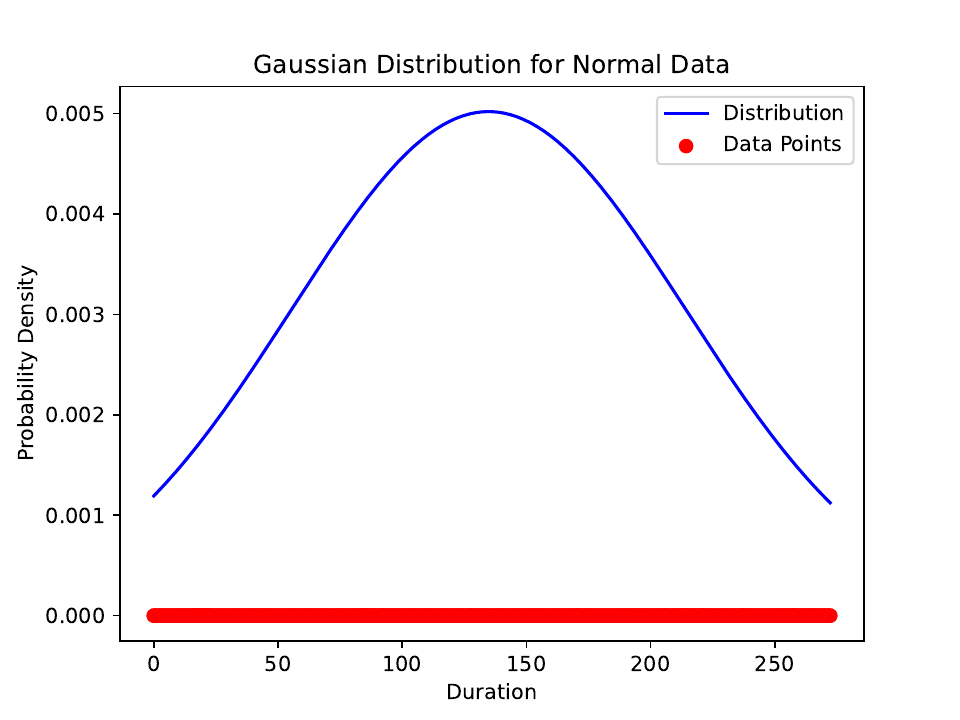}}  
    
    \subfigure[Probability distribution for attack traffic\label{attack}]{\includegraphics[width=.4\textwidth]{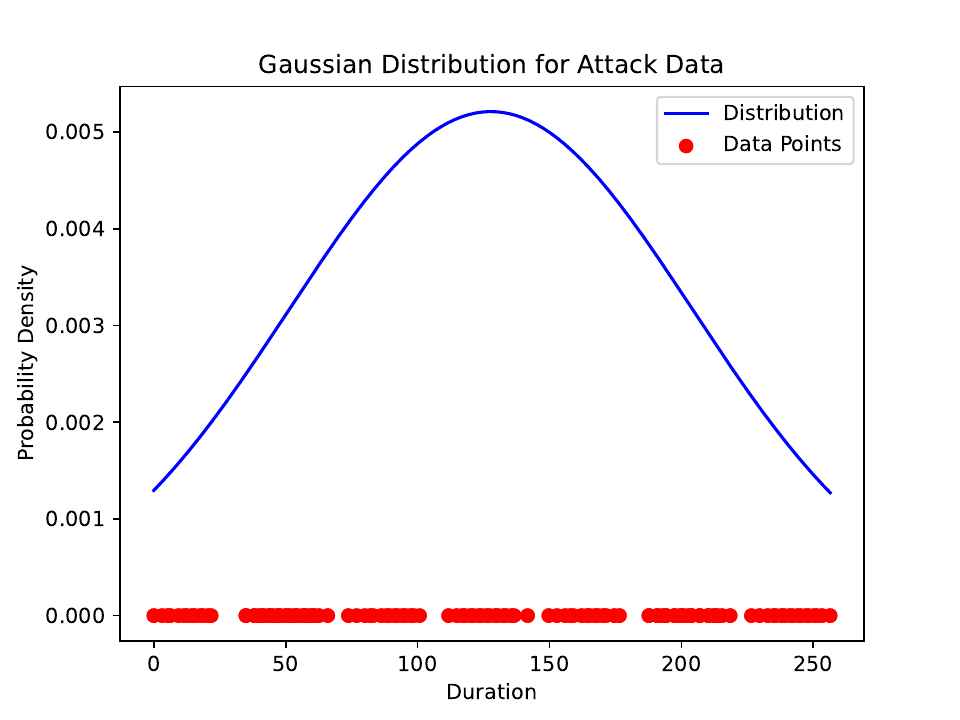}}
    \caption{Probability Distribution}    
\end{figure}

To understand the pattern of legitimate and attack packets, probability density curves are plotted, as shown in Fig. \ref{normal} and Fig. \ref{attack}. It is observed that both legitimate and attack traffic flows follow Gaussian distribution. However, there is a clear difference in packet arrival frequency between both cases.

\begin{figure*}[ht!]
    \centering
    \subfigure[\textcolor{black}{Comparison of duration of attack and normal traffic} \label{a}]{\includegraphics[width=0.3\textwidth]{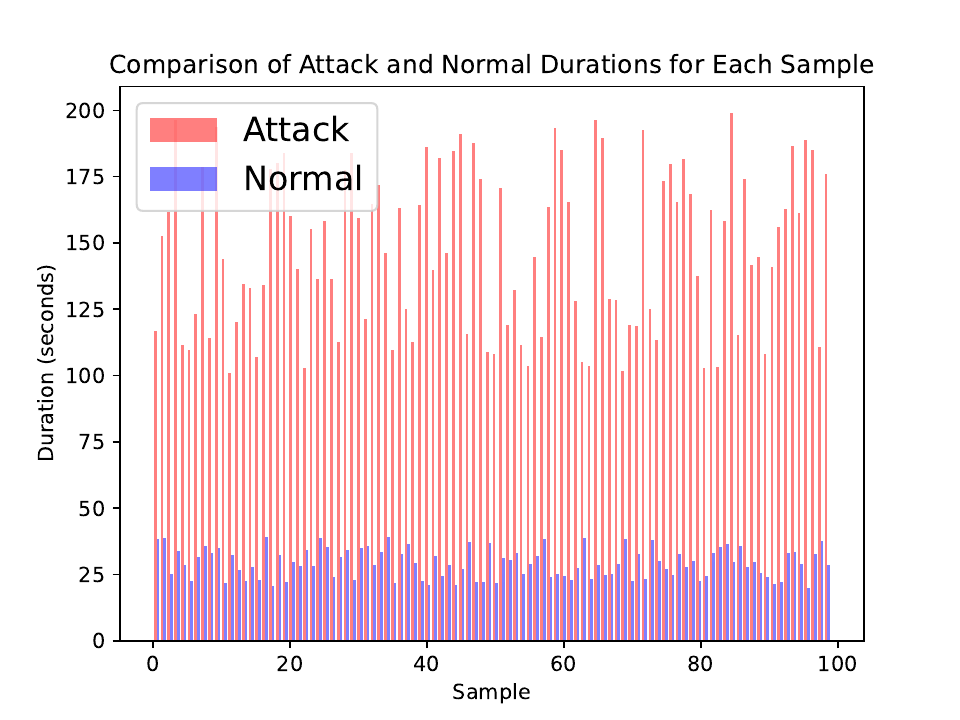}}
        \subfigure[\textcolor{black}{Comparison of packets of attack and normal traffic}\label{b}]{\includegraphics[width=0.3\textwidth]{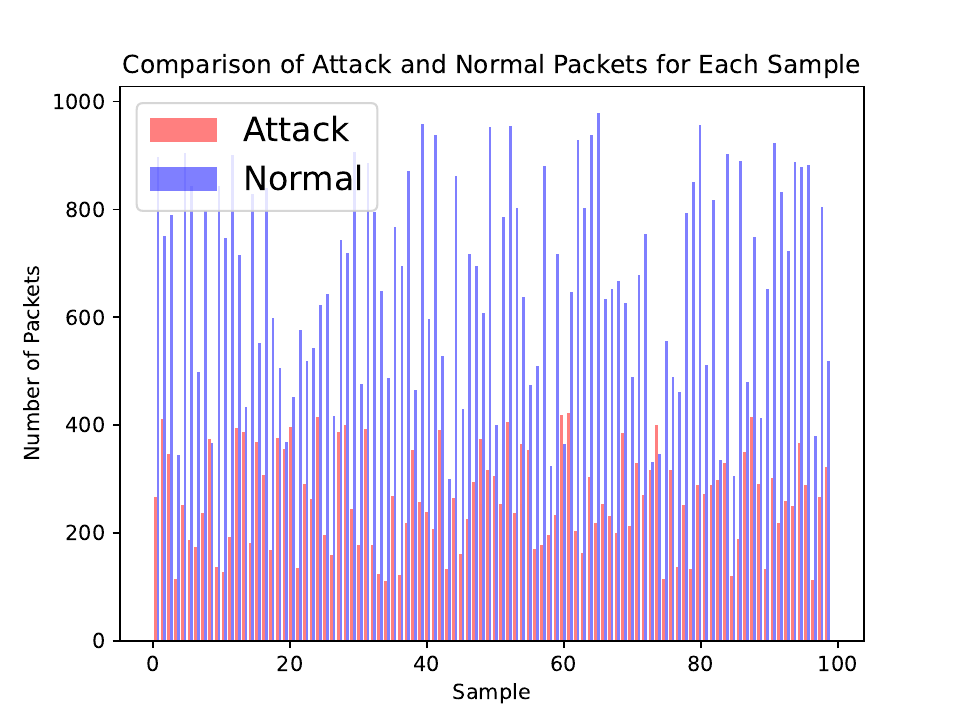}}          
    \subfigure[\textcolor{black}{Comparison of bytes of attack and normal traffic}\label{c}]{\includegraphics[width=0.3\textwidth]{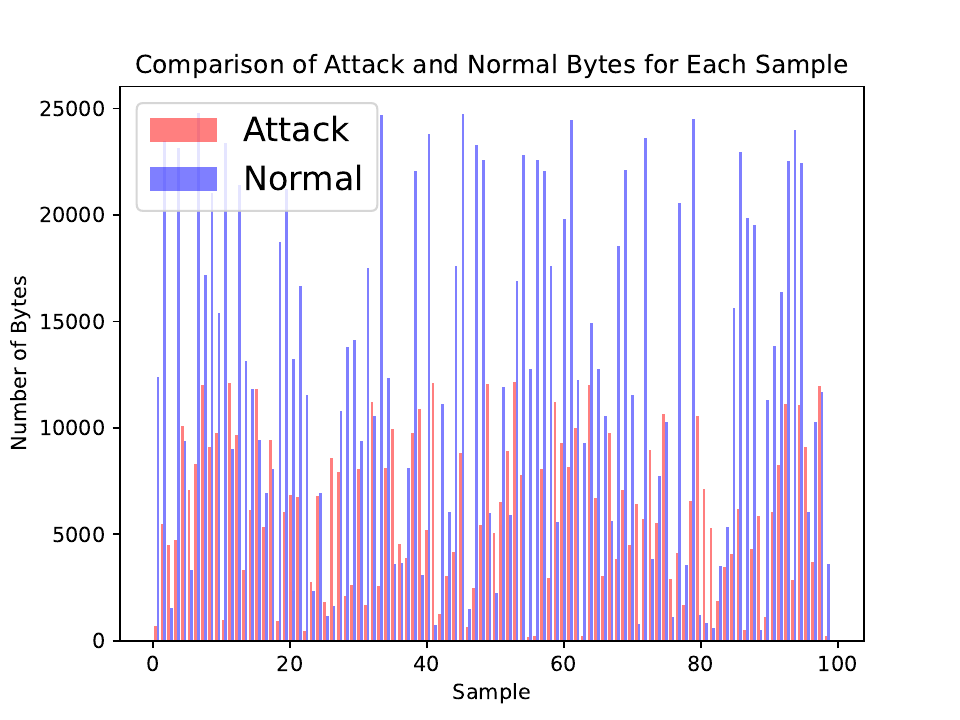}}
        \caption{\textcolor{black}{Comparison of duration, packets and bytes}}
    \label{dpb}
    \end{figure*}
\begin{figure*}[ht!]
   \subfigure[\textcolor{black}{Comparison of mean duration of attack and normal traffic}\label{md}]{\includegraphics[width=0.3\textwidth]{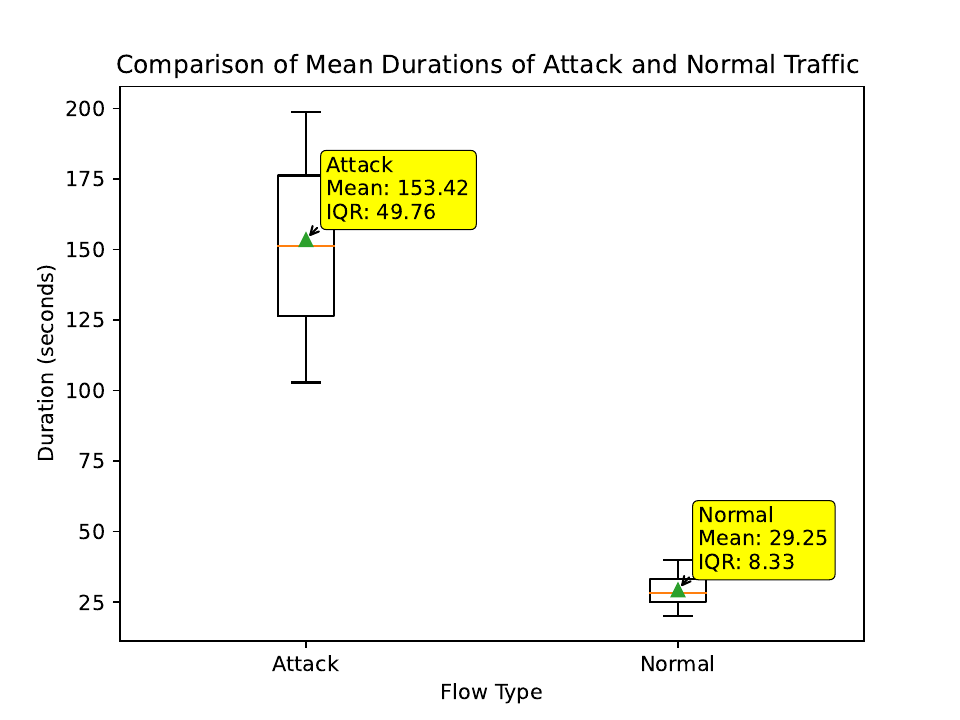}}
    \subfigure[\textcolor{black}{Comparison of mean packets of attack and normal traffic}\label{mp}]{\includegraphics[width=0.3\textwidth]{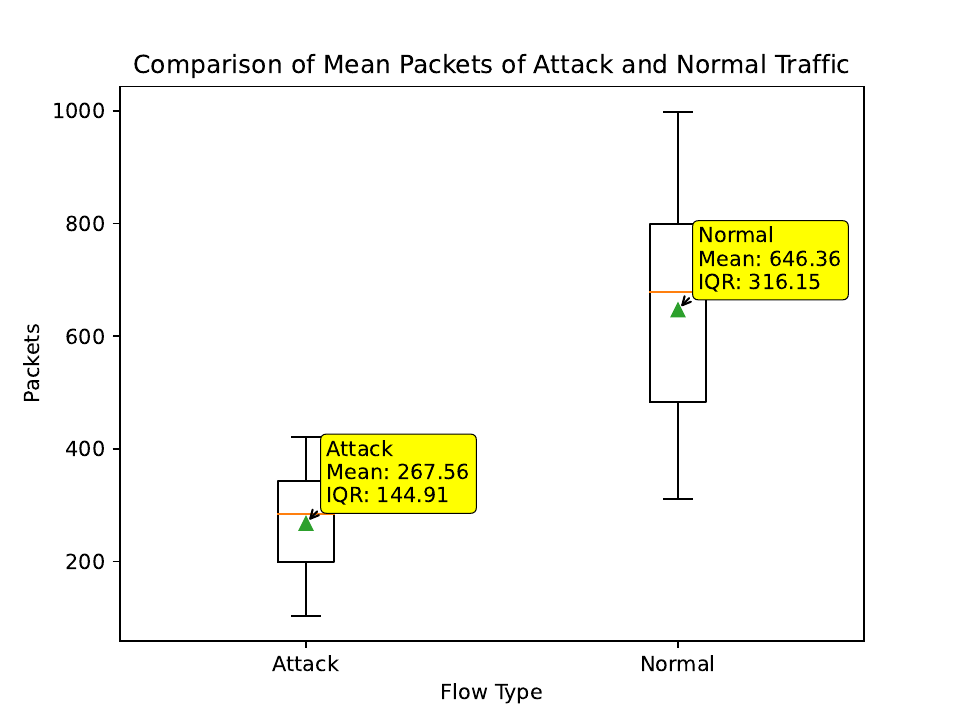}}     
    \subfigure[\textcolor{black}{Comparison of mean bytes of attack and normal traffic}\label{mb}]{  \includegraphics[width=0.3\textwidth]{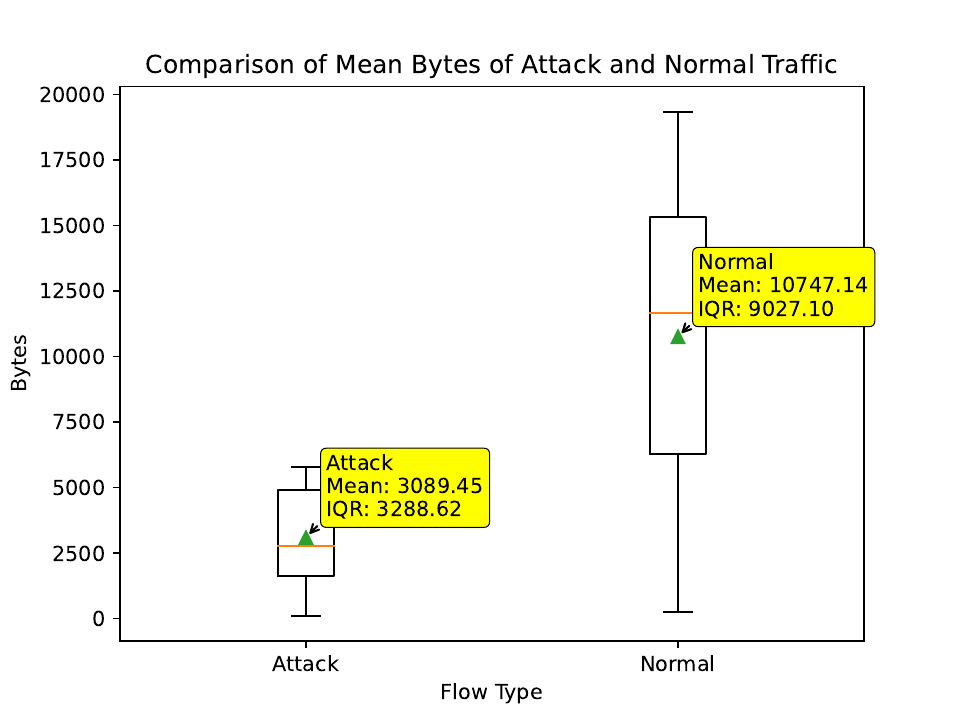}}      
        \caption{\textcolor{black}{Comparison of mean duration, mean packets and mean bytes}}
    \label{mdpb}
    \end{figure*}
    
\begin{figure}
    \centering
    \subfigure[Mean packet arrival frequency \label{paf}]{\includegraphics[width=.4\textwidth]{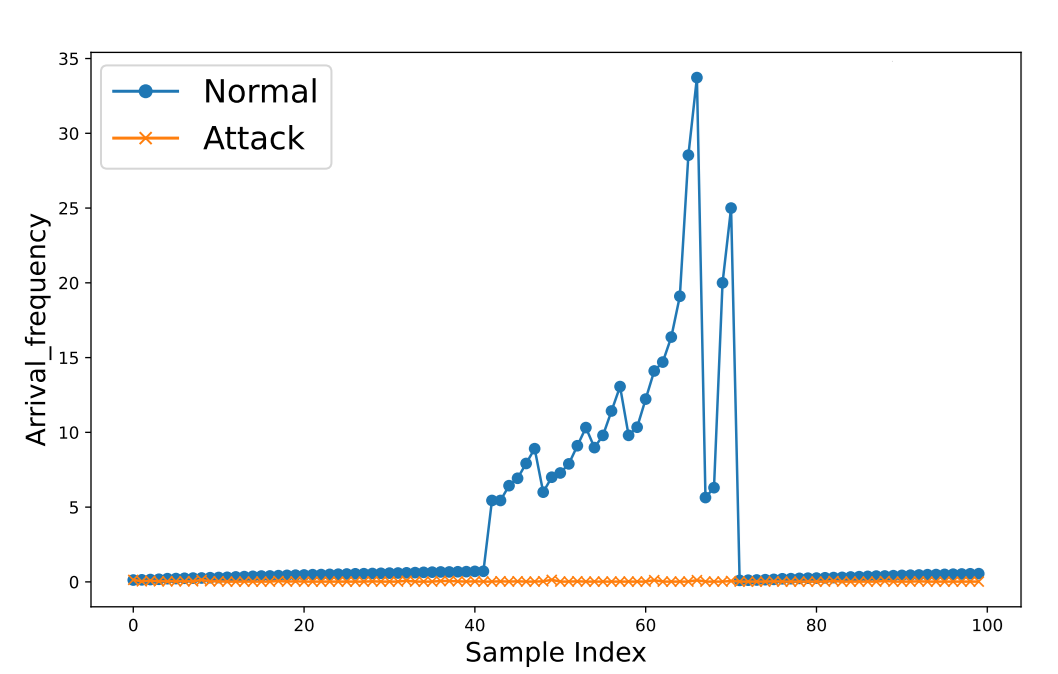}}
    \subfigure[Mean information gain\label{crs}]{\includegraphics[width=.4\textwidth]{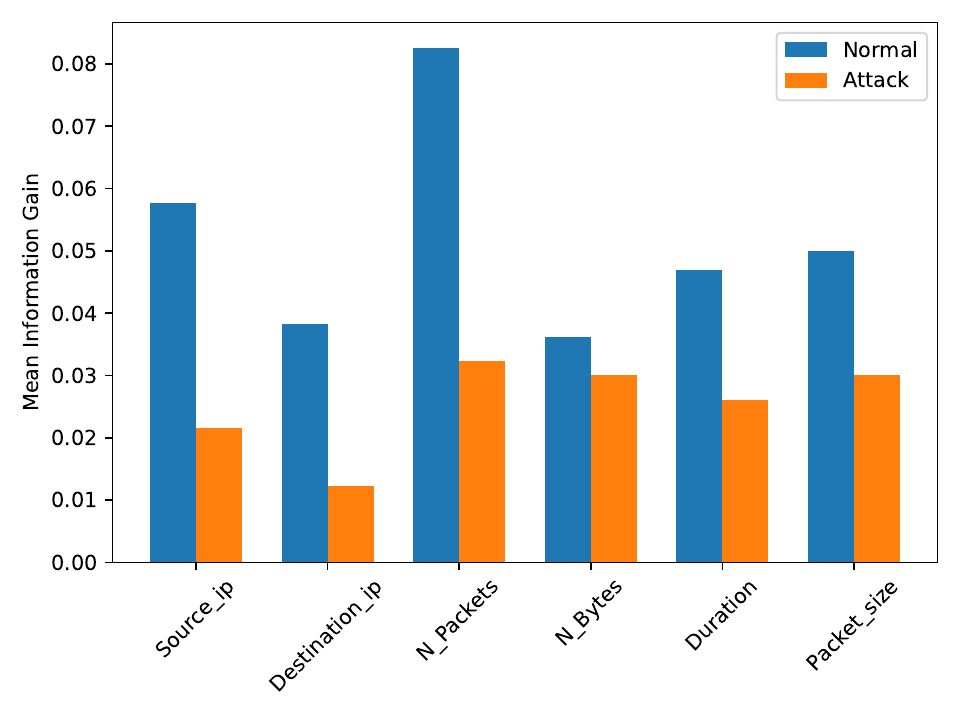}}
    \caption{PAF and IG}
\end{figure}

This module performs an analysis to understand the traffic pattern and assigns it a value of 0 (normal) or 1 (attack). The comparative analysis of attack and normal flow is given in Fig. \ref{paf}.

\subsubsection{Content Relevance Score (CRS)}
This module calculates the usefulness of the packet based on various features taken from packet content. To detect any possible anomaly, every packet plays an important role because each packet contains some meaningful information that helps in finding the difference between normal and potentially malicious traffic. Information gain \cite{lei2012feature} for various attributes of normal flow and attack flow is shown in Fig. \ref{crs}. To calculate the usefulness of the packet, information gain is a good metric \cite{cheng2023pcmigr}. Source IP, destination IP, packet size, payload size, entropy of payload content, and duration are the features for calculating usefulness. For each packet in a flow, the entropy of the above-mentioned features is calculated using Shannon's entropy, which is mathematically expressed as Eq. \ref{H}.
    \begin{figure*}[ht!]
         \subfigure[Comparison of coefficient of variance of duration of attack and normal traffic\label{cvd}]{ \includegraphics[width=0.3\textwidth]{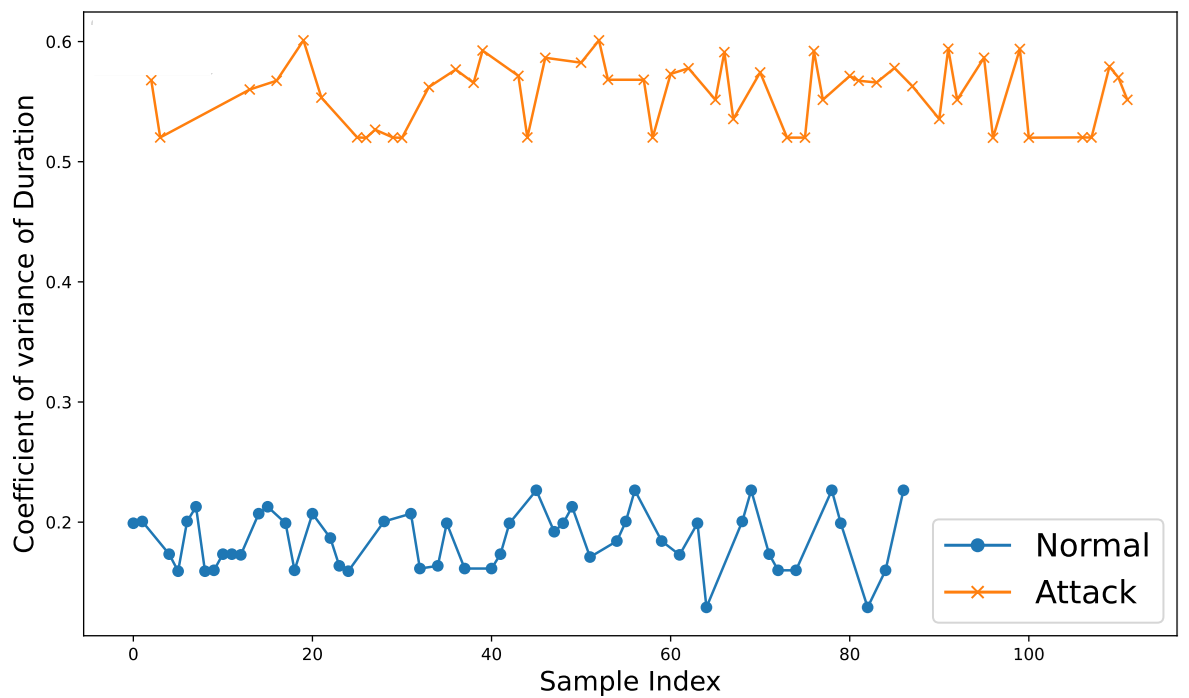}}
           \subfigure[Comparison of coefficient of variance of packets of attack and normal traffic\label{cvp}]{\includegraphics[width=0.3\textwidth]{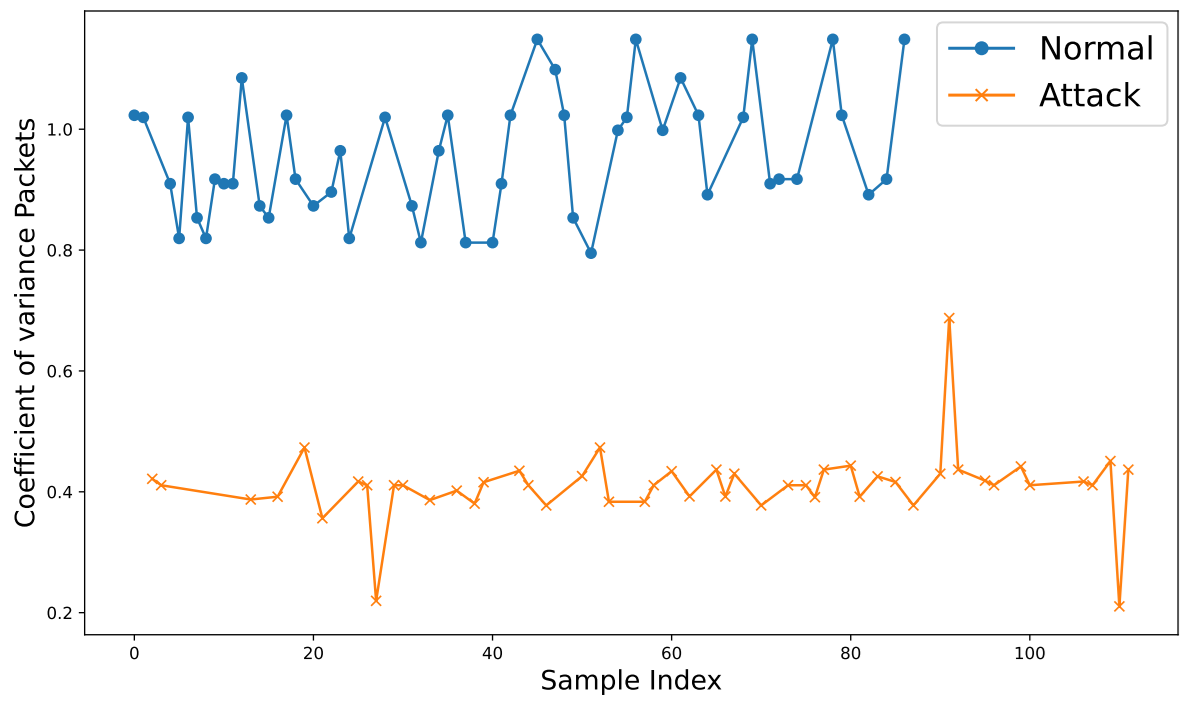}}
    \subfigure[Comparison of coefficient of variance of duration of attack and normal traffic\label{cvb}]{\includegraphics[width=0.3\textwidth]{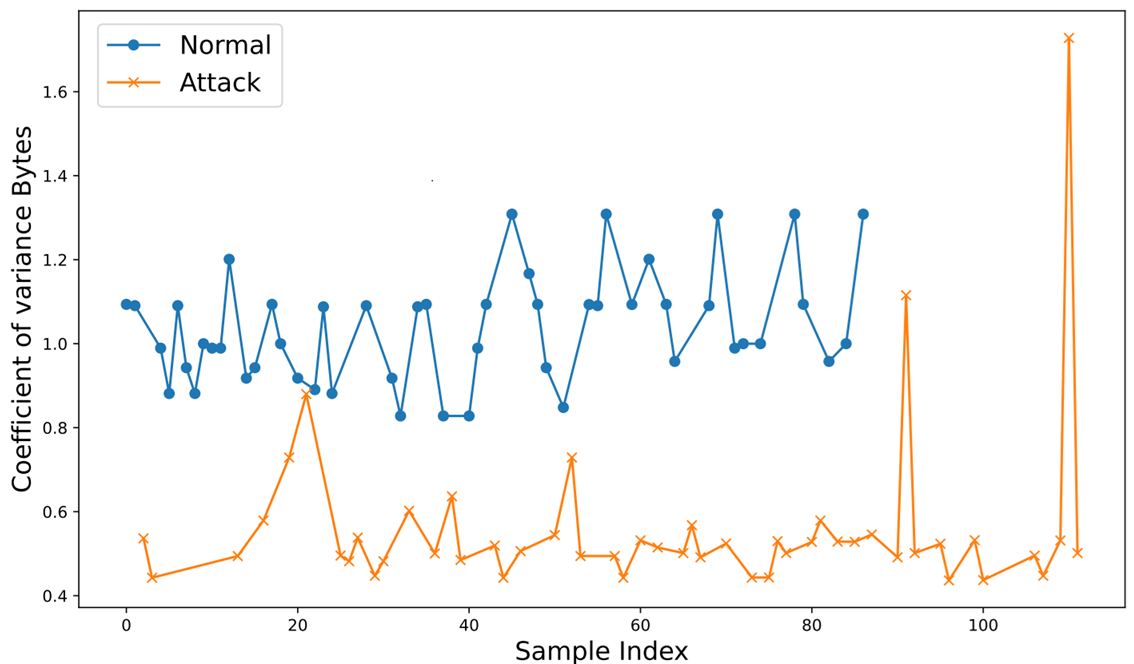}}
        \caption{Comparison of CVD, CVP and CVB}
    \label{fig:multiple_diagrams}
\end{figure*}
\begin{equation}
    H(D) = - \sum_{i=1}^n P(D_i) \cdot \log_2(P(D_i))
    \label{H}
\end{equation}

After calculating the entropy, cumulative information gain for all features is calculated and assigned to the packets using Eq. \ref{ig}. 

\begin{equation}
IG(A) = H(D)-\sum_{i=1}^n \frac{|D_i|}{|D|}\cdot H(D_i)
\label{ig}
\end{equation}
where, $IG$ represents information gain, and $A$ is the attribute for which $IG$ is to be calculated, $D$ represents the size of the dataset, $D_i$ represents $i^{\text{th}}$ attribute of dataset, $H$ is the entropy.
\subsubsection{Identification of possible spoofed IP}
To launch a low-rate attack, The attacker needs many compromised hosts. Due to the difficulty in compromising the required number of hosts, there is a possibility that the attacker may use spoofed IP to launch the attack. FloRa uses BCP38 \cite{cao2018understanding} to detect spoofed IP. BCP38 is an ingress filtering technique that detects spoofed IPs to avoid DDoS attacks. It enforces source IP validation at network ingress points and maintains a list of IP addresses allocated within the network. When a packet arrives, BCP38 ensures that the IP of the incoming packet matches with the maintained list. In FloRa, BCP38 is implemented in the anomaly predictor module, where it compares the list of IPs provided by CRS with the maintained list of IPs to check the possibility of spoofed IPs.

\subsection{Feature Selector}
In this module, the essential features are selected to make classification, through which attack could be detected. Three new features, Packet Arrival Frequency (PAF), Content Relevance Score (CRS), and Possible Spoofed IP (PSI), are introduced in FloRa. Similarly, other features such as duration, number of bytes, and number of packets indicate significant differences between flow tables under attack traffic and legitimate traffic. The duration indicates the lifetime of the flow rule, the number of bytes represents the total number of matches with the corresponding flow rule, and the number of packets represents the total packets matched with the corresponding flow. The basic idea behind the attack is to consume the storage capacity of the flow table and keep it occupied for a longer duration. To achieve this, the attack flow follows a pattern, due to which there will be a significant change in the distribution of duration, bytes and packets, their mean and coefficient of variation as shown in Fig. \ref{dpb}, Fig. \ref{mdpb} and Fig. \ref{fig:multiple_diagrams}.

\begin{figure}[h]
    \centering
    \includegraphics[width=.45\textwidth]{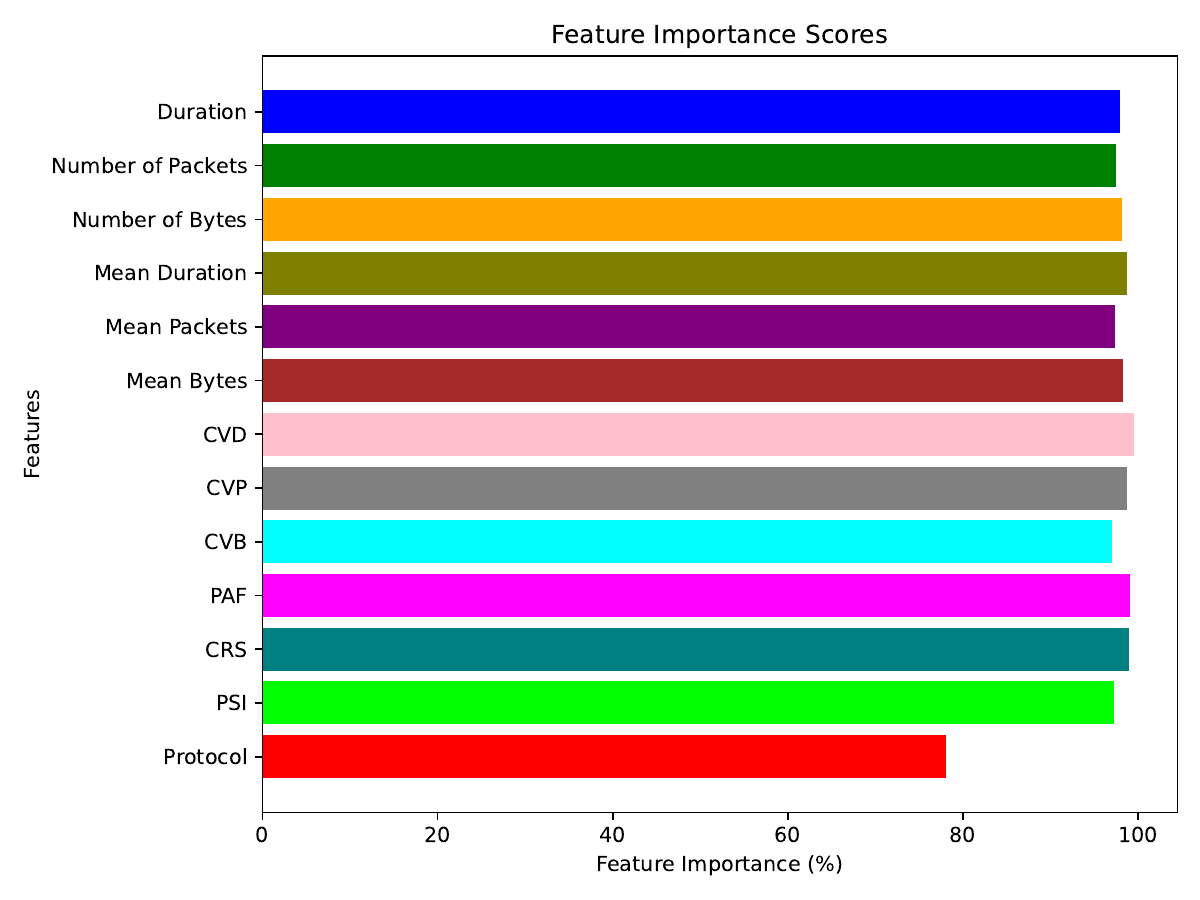}
    \caption{Feature importance scores}
    \label{fis}
\end{figure}

Along with the above-mentioned features, some other features are also considered in FloRa. The importance of all the above-mentioned features is computed using Recursive feature elimination with cross-validation (RFECV) \cite{zeleke2021efficient}, which is presented in Fig. \ref{fis}. The detection dataset includes 12 most important features with the highest score.

\subsection{Attack Detection}
Cat Boost \cite{ibrahim2020comparison}, which is employed as the attack detection algorithm, is a powerful gradient-boosting technique. Cat Boost overcomes the limitations of other gradient boosting algorithms such as LightGBM \cite{ke2017lightgbm} and XGBoost \cite{sun2021novel} due to its specialization in categorical features. As an advancement among other algorithms, Cat Boost can handle categorical features without using one-hot encoding \cite{cerda2020encoding}, making it efficient in pre-processing efforts and dimensionality-related issues. To address the problem of over-fitting, Cat Boost uses a combination of ordered boosting and oblivious trees. Another critical attribute of Cat Boost is ``exclusive feature building (EFB)", which correlates grouped features to address dimensionality-related issues, resulting in improved training speed. Along with the above-discussed features, it improves a leaf-wise tree development strategy with depth restriction, which helps reduce computational overhead and makes Cat Boost suitable for large-scale and real-world datasets. Algorithm \ref{detection} states the steps of detection. Further explanations are presented in Section \ref{ee}.
% \begin{algorithm}
% \caption{Pseudocode of Detection method}\label{detection}
% \begin{algorithmic}[1]
% \State $R$ - [...] \Comment{List containing flow rules} 
% \State $D$ - [...] \Comment{List containing flow rule duration}
% \State $Dataset$ - [\,] \Comment{Empty list for dataset}
% \State $Blacklist$ - [\,] \Comment{Empty list of blacklisted IP}

% \For{each $i$ in $R$}
%     \If{$i > 100$}
%         \State $S \leftarrow$ [\,]
%         \For{each $j$ in $D$}
%             \If{$j > 100$}
%                 \State Add $j$ to $S$
%             \EndIf
%         \EndFor
        
%         \State $PAF\_score \leftarrow$ calculate\_PAF($S$)
%         \State $CRS \leftarrow$ calculate\_CRS($S$)
%         \State $PSI \leftarrow$ BCP38($S$)

%         \For{each flow in the dataset}
%             \If{$PAF\_score < t_\text{idle}$ or $CRS < \text{50\%\,}$ or $PSI = \text{True}$}
%                 \State $Dataset[\,].append(flow)$
%             \EndIf
%         \EndFor
%     \EndIf
% \EndFor

% \For{each $i$ in $Dataset$}
%     \If{Cat Boost($i$).label $= 1$}
%         \State Add $Blacklist[\,].append (i)$
%     \Else
%         \State Continue
%     \EndIf
% \EndFor
% \end{algorithmic}
% \end{algorithm}
\begin{algorithm}
\caption{Pseudocode of Detection method}\label{detection}
\begin{algorithmic}[1]
\State $R$ - \Comment{List containing flow rules}
\State $D$ - \Comment{List containing flow rule duration}
\State $Dataset$ - [] \Comment{Empty list for dataset}
\State $Blacklist$ - [] \Comment{Empty list of blacklisted IP}

\For{each $i$ in $R$}
    \If{$i > 100$}
        \State $S \leftarrow$ [] \Comment{Create an empty list $S$}
        \For{each $j$ in $D$}
            \If{$j > 100$}
                \State Add $j$ to $S$
            \EndIf
        \EndFor

        \State $PAF\_score \leftarrow$ calculate\_PAF($S$)
        \State $CRS \leftarrow$ calculate\_CRS($S$)
        \State $PSI \leftarrow$ BCP38($S$)

        \For{each flow in the dataset}
            \If{$PAF\_score < t_\text{idle}$ or $CRS < \text{50\%\,}$ or $PSI = \text{True}$}
                \State $Dataset[\,].append(flow)$
            \EndIf
        \EndFor
    \EndIf
\EndFor

\For{each $i$ in $Dataset$}
    \If{Cat Boost($i$).label $= 1$}
        \State Add $Blacklist[\,].append (i)$
    \Else
        \State Continue
    \EndIf
\EndFor
\end{algorithmic}
\end{algorithm}

In the detection module, the model is trained using a pre-processed dataset in five sets. In the first set, the (training, testing) size is (80\%, 20\%). The second set  (training, testing) size is (75\%, 25\%,). The consecutive size are (70\%, 30\%), (65\%, 35\%), and (60\%, 40\%).
The trained model is then deployed over the network for classification. It is observed that the model achieves detection accuracy of 99.49\%\,. 
\subsection{Attack Mitigation}
\textcolor{black}{Upon detection of the LOFT attack, the mitigation process is initiated by FloRa in three sequential steps. Initially, FloRa scrutinizes each flow rule, extracting six distinct features and employing Cat Boost for classification to discern between normal and attack flow rules. Identified attack flow rules are then promptly added to the eviction list. Subsequently, FloRa orchestrates the systematic removal of these malicious flow rules from the switch's flow table, thereby liberating space for legitimate traffic. Furthermore, to safeguard essential elephant flows integral to network operations, FloRa meticulously filters them to prevent inadvertent deletion. Concurrently, FloRa maintains a blacklist of suspicious source IPs, tracking their deletion frequency. IPs surpassing predefined thresholds are expeditiously blocked, ensuring the network's integrity and stability. The controller further blocks the corresponding IPs to stop the attack flow. This comprehensive mitigation strategy ensures efficient resource utilization while thwarting potential threats posed by the LOFT attack.}

\section{Experiment and evaluation}\label{ee}
\subsection{Experimental setup}
\begin{figure}[h]
    \centering
    \includegraphics[width=.5\textwidth]{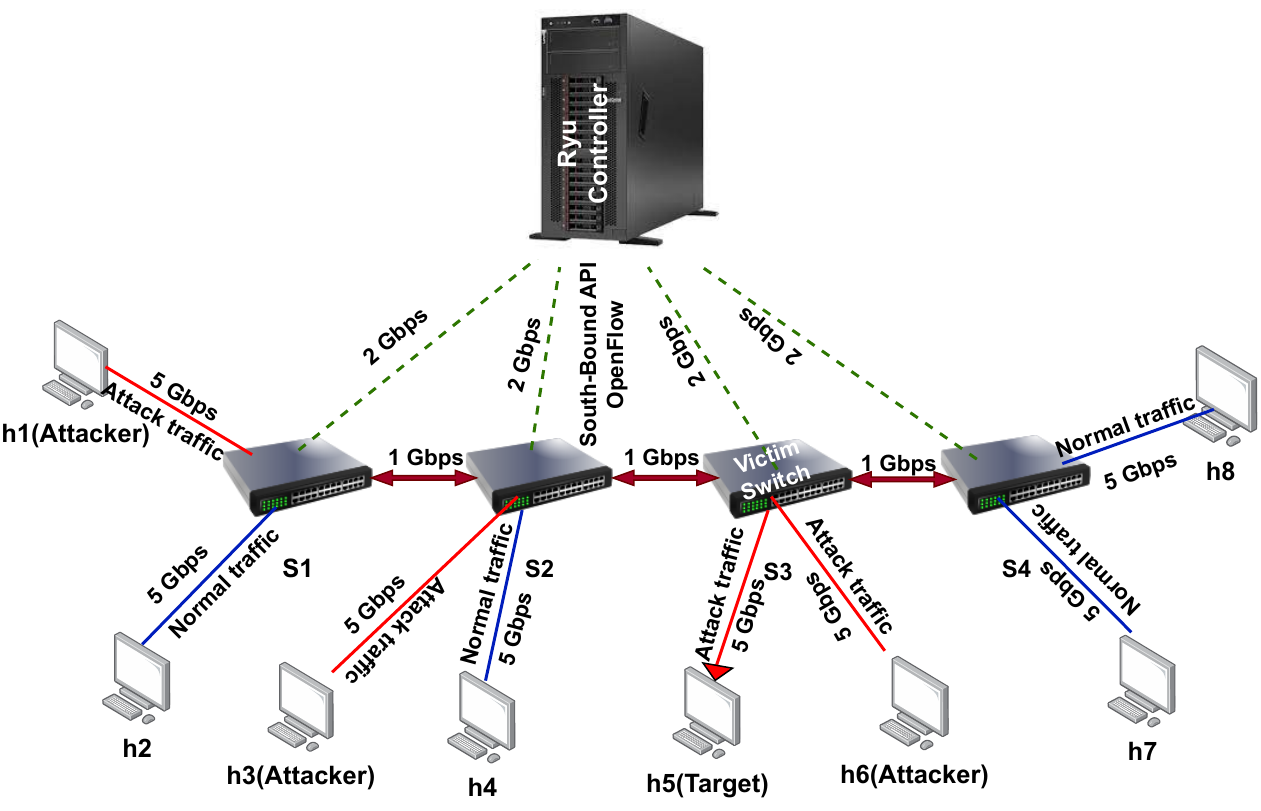}
    \caption{Experimental Topology}
    %\mb{seems very simple topology and diagram while you are talking about SDN security. So I suggest to revise it}}
    \label{topo}
\end{figure}

For the experiment, Ryu (version 4.3.4) and Mininet (version 2.3.0) are used to simulate the SDN over the Ubuntu 16.04 LTS operating system. The experimental setup uses OF version 1.5.0 with match fields such as source IP, destination IP, source port, destination port, etc. Mininet tree topology is used for the experiment depicted in Fig. \ref{topo}. The topology includes one SDN controller, four OF switches, and eight user hosts, out of which h1, h3, and h6 are the attackers. The link bandwidth between switches is 1 Gbps, and the link bandwidth between switch and host is 5 Gbps. The OF switch is configured with a FIFO replacement policy with an idle timeout of 20 seconds.

\subsection{Dataset}
\subsubsection{Background flow}
\textcolor{black}{We employ the tcpreplay tool \cite{tcpreplay} to reproduce the IMC-10 data center network trace \cite{benson2010data} for simulating SDN network flows. Specifically, we select pt2 from univ1 as representative of normal background flows. Considering that this network traffic dataset was collected approximately ten years ago, we opt for flow table sizes of $1500$, $2000$, $3000$ and $4000$ to better align with the hardware capacities available at that time \cite{yang2019stereos}. Table \ref{tab:bts} presents the parameters for configuring the background traffic.} 
% IMC-10 network trace \cite{benson2010data} simulates network traffic using \cite{tcpreplay}. The pt2 of univ1 is used to simulate the normal background flow. Table \ref{tab:bts} shows the setting for simulating the background traffic. The attack is carried out in four sets with different flow table capacities and packet transmission rates. The flow tables are collected every second. Each set of attacks lasts for $1000$ second, the attack starts at $300^{th}$ second. A total number of $19715$ legitimate flows and $20235$ attack flows are collected and added to the detection dataset after labelling.
\subsubsection{Attack execution}
\textcolor{black}{To execute the LOFT attack, the Scapy \cite{rohith2018scapy} library is employed. The attack is executed in 4 sets, details as mentioned in Table \ref{tab:as}. While implementing the attack, ANP must be moderate. Similarly, AF should be set so that the existing attack flow rules are not deleted. The packet size of attack packets is set after analyzing the background traffic within the range of the normal flow packets.}
\begin{table}[ht]
\centering
\renewcommand{\arraystretch}{1.5} % Increase the vertical spacing by 50%
\caption{Background traffic setting}
\begin{tabular}{|c|c|c|}
\hline
S. No. & Capacity (entries) & Transmission Rate (pps) \\
\hline
1. & 1500 & 250\\
\hline
2. & 2000 & 300\\
\hline
3. & 2500 & 400\\
\hline
4. & 3000 & 600\\
\hline
\end{tabular}
\label{tab:bts}
\end{table}
\subsubsection{Detection dataset}
\textcolor{black}{The attack is carried out in four sets with different flow table capacities and packet transmission rates. The flow tables are collected every second. Each set lasts for 1000 second, the attack starts at $300^th$ second. A total number of $39950$ flows are collected out of which $19715$ are legitimate flows and $20235$ are attack flows. The collected flows are added to the detection dataset after labelling.}

% \subsection{Dataset generation}
% IMC-10 network trace \cite{benson2010data} simulates network traffic using \cite{tcpreplay}. The pt2 of univ1 is used to simulate the normal background flow. Table \ref{tab:bts} shows the setting for simulating the background traffic. The attack is carried out in four sets with different flow table capacities and packet transmission rates. The flow tables are collected every second. Each set of attacks lasts for $1000$ second, the attack starts at $300^{th}$ second. A total number of $19715$ legitimate flows and $20235$ attack flows are collected and added to the detection dataset after labelling.

\begin{table}[ht]
\centering
\renewcommand{\arraystretch}{1.5} % Increase the vertical spacing by 50%
\caption{Attack Setting}
\begin{tabular}{|c|c|c|}
\hline
S. No. & AF (s) & ANP (pkt/AF) \\
\hline
1. & (4,8) & 20 \\
\hline
2. & (8,12) & 40 \\
\hline
3. & (12,16) & 60 \\
\hline
4. & (16,19) & 80 \\
\hline
\end{tabular}
\label{tab:as}
\end{table}

\subsection{Classification Effectiveness}
FloRa aims to achieve better accuracy and low FPR and FNR detection. Table \ref{Tab:comp1} shows the performance comparison of FloRa with existing methods.

\begin{table}[h]
\centering
\renewcommand{\arraystretch}{1.5} % Increase the vertical spacing by 50%
\caption{Detection performance comparison with existing methods}
\label{tab:classifier-metrics}
\begin{tabular}{|l|c|c|c|c|}
\hline
\textbf{Method} & \textbf{Accuracy (Acc)} & \textbf{FPR} & \textbf{FNR} & \textbf{F1-Score} \\
\hline
ECSD \cite{dinh2020ecsd} & 96.99\% & 1.02\% & 3.96\% & 97.05\% \\
\hline
FTMaster\cite{tang2023ftmaster} & 99.33\% & 0.03\% & 1.29\% & 99.33\% \\
\hline
SFTOGaurd \cite{tang2023sfto} & 98.75\% & 1.13\% & 2.25\% & 98.73\% \\
\hline
ARIMA\cite{dinh2019abnormal} & 94.88\% & 3.22\% & 3.84\% & 95.01\% \\
\hline
FloRa & 99.49\% & 0.01\% & 1.22\% & 99.87\% \\
\hline
\end{tabular}
\label{Tab:comp1}
\end{table}

 FloRa attains an accuracy of 99.49\%\, FPR 0.01\%\, FNR 1.22\%\, and F-1 Score 99.87\%\, which is better than the existing methods. FloRa exhibits superior performance across various evaluation metrics when compared to other methods. Its high accuracy, low false positive rate, low false negative rate, and impressive F1-Score make it a compelling choice for detecting LOFT attacks. FloRa outperforms its counterparts in terms of effectiveness and reliability. FloRa ensures reduced CPU and Memory overhead in comparison with existing detection schemes. The classification rate of FloRa is also better than that of existing methods. FloRa employs Cat Boost as a detection classifier to detect low-rate DDoS attacks. To justify the merits of Cat Boost, its performance is compared with five other classifiers, XGBoost, Light GBM, Random Forest, Logistic Regression, and Adaboost, which are applied to our created dataset. Table \ref{Tab:comp2} shows that the performance of Cat Boost is better than all other mentioned classifiers. 

 \textcolor{black}{Additionally, performance of various methods employed in Software-Defined Networking (SDN) environments, including RIT, LOFTGUARD, and Sphinix are also compared with FloRa as presented in Table \ref{tab:loft_attacks}. We examined key performance metrics such as mean utilization, total overflows, CPU usage, and memory consumption. Notably, FloRa emerged as the standout performer, demonstrating a balanced utilization rate, minimal total overflow incidents, and efficient resource utilization. With a mean utilization of 62.23\%, FloRa maintains stability while exhibiting CPU usage of 1.86\% and requiring moderate memory of 217 MB. In contrast, RIT achieves a higher mean utilization of 83.98\% and suffers from a significantly higher total overflows of 49. LOFTGUARD utilizes 89.58\% of flow table, albeit with 17 number of total overflows and higher CPU usage of 2.11\% and memory consumption of 178 MB. Sphinix, while achieving a mean utilization of 84.85\%, experiences 36 total overflows with moderate CPU usage of 1.14\% and memory requirement of 49 MB. Overall, FloRa's balanced performance across various metrics positions it as the optimal choice for stable and reliable method in SDN environments.}

\begin{figure}
    \centering
    \subfigure[ Flow Table size 1500\label{wf1}]{\includegraphics[width=0.45\textwidth]{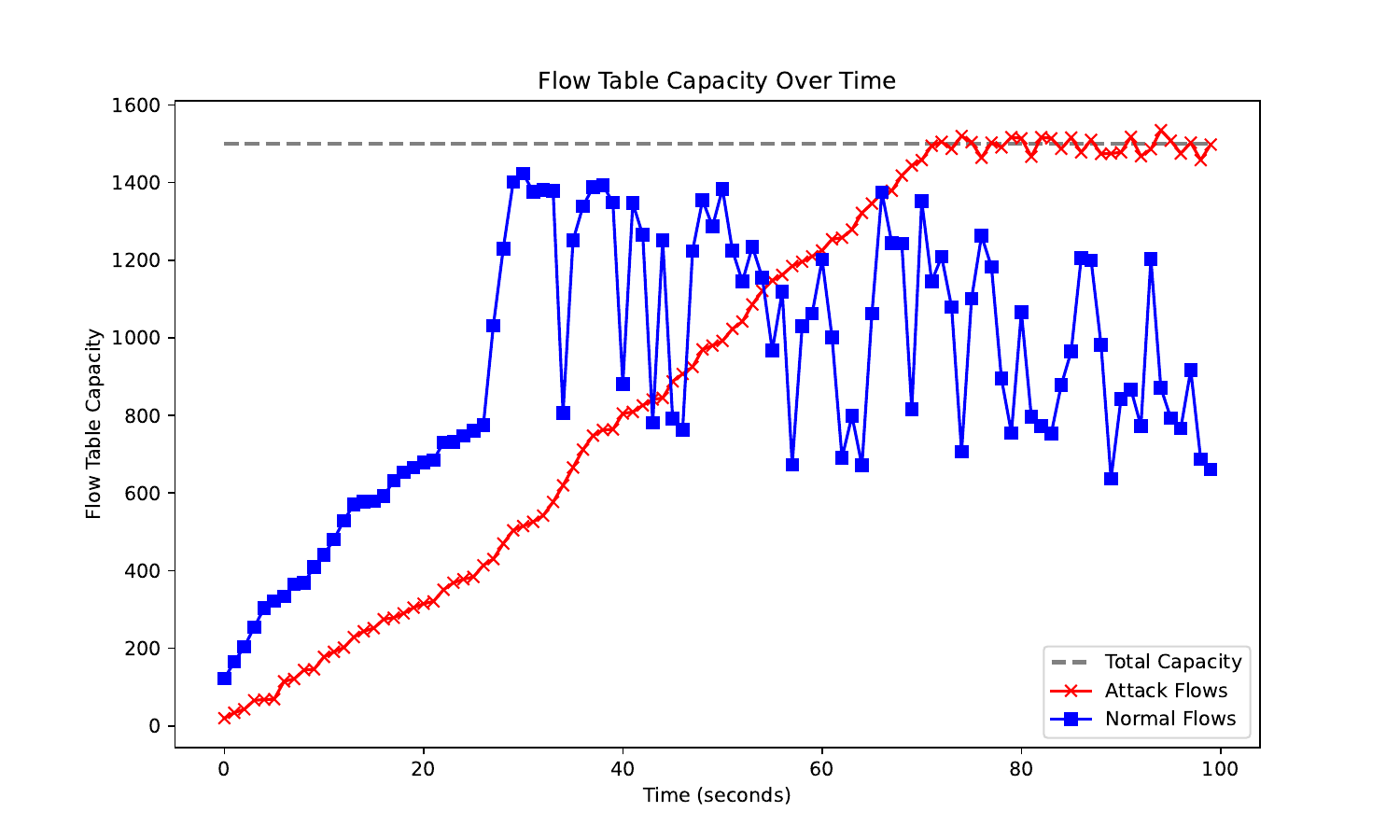}}
    
    \subfigure[Flow Table size 2000\label{wf2}]{\includegraphics[width=0.45\textwidth]{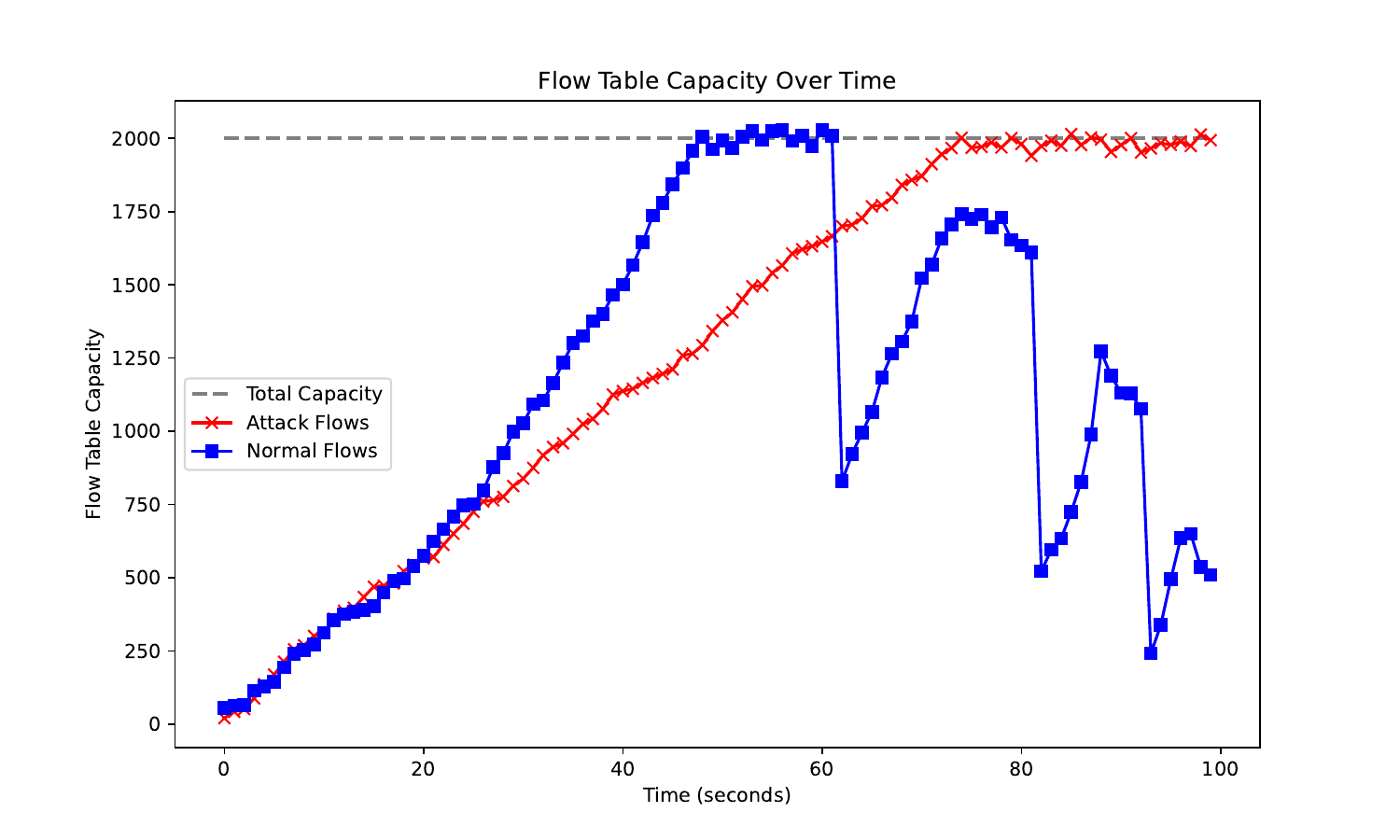}}
            
    \subfigure[Flow Table size 2500\label{wf3}]{\includegraphics[width=0.45\textwidth]{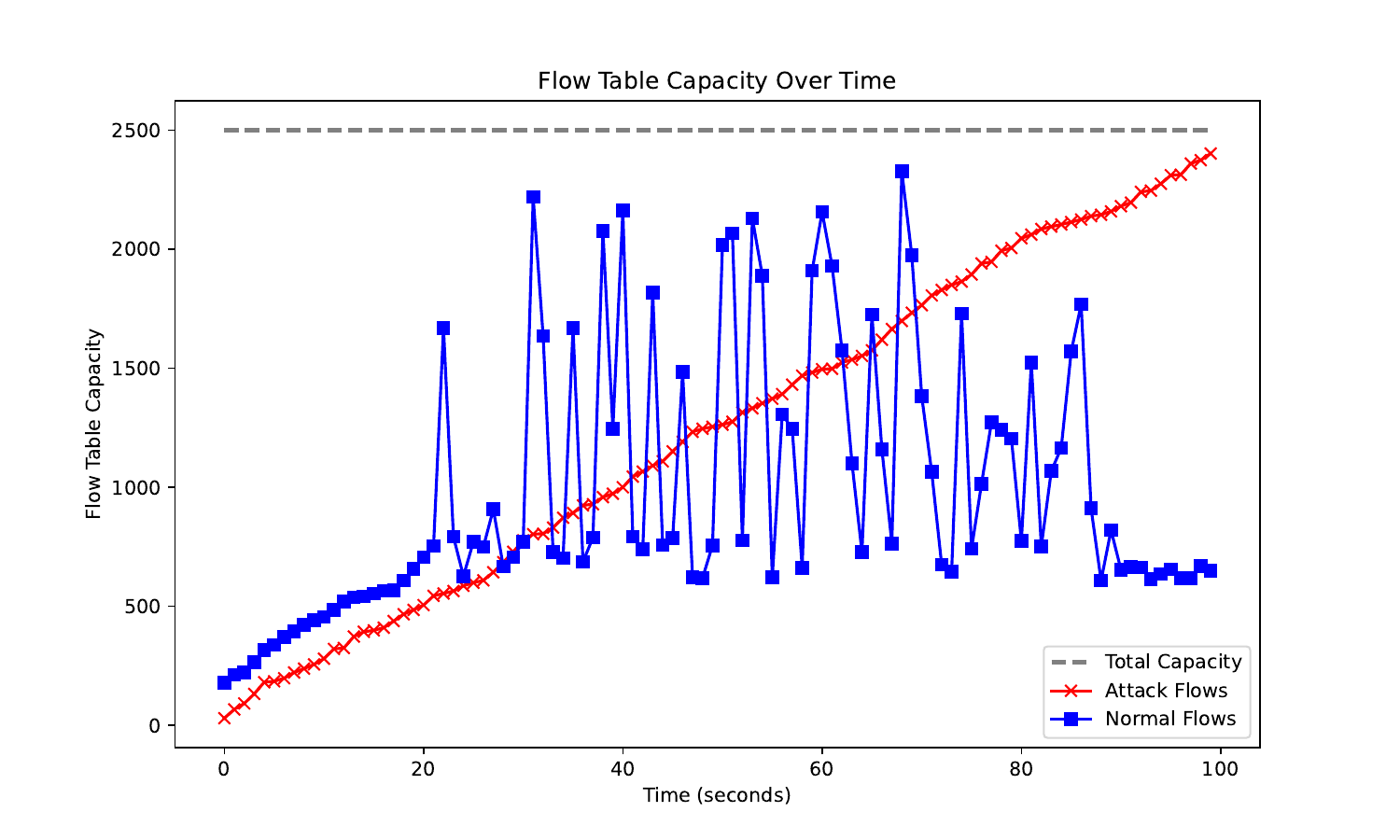}}
    \subfigure[Flow Table size 3000\label{wf4}]{\includegraphics[width=0.45\textwidth]{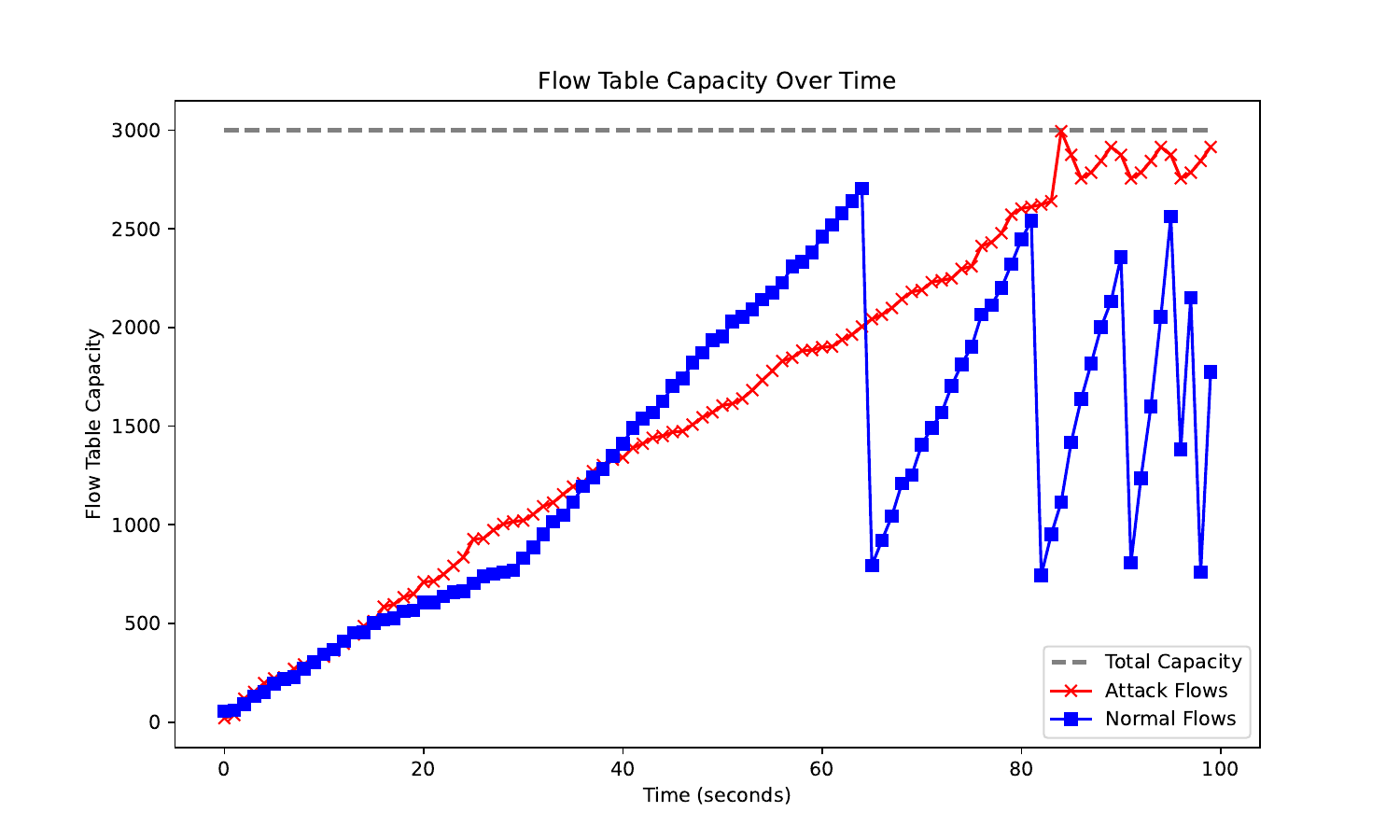}}
    
    \caption{Normal and attack flow rules without FloRa}
    \label{WF}
    \end{figure}

\begin{table}[h]
\centering
\renewcommand{\arraystretch}{1.5} % Increase the vertical spacing by 50%
\caption{Classifier Performance Metrics}
\label{tab:classifier-metrics}
\begin{tabular}{|l|c|c|c|c|}
\hline
\textbf{Classifier} & \textbf{Acc} & \textbf{FPR} & \textbf{FNR} & \textbf{F1-Score} \\
\hline
Light GBM & 98.33\% & 1.30\% & 2.87\% & 98.28\% \\
\hline
XGBoost & 98.57\% & 1.03\% & 2.29\% & 98.61\% \\
\hline
AdaBoost & 98.46\% & 1.22\% & 1.84\% & 98.46\% \\
\hline
Logistic Regression & 96.02\% & 2.18\% & 3.62\% & 96.07\% \\
\hline
Random Forest & 98.35\% & 2.11\% & 1.25\% & 99.32\% \\
\hline
Cat Boost  & 99.49\% & 0.01\% & 1.22\% & 99.87\% \\
\hline
\end{tabular}
\label{Tab:comp2}
\end{table}
\begin{table}[h]
  \centering
  \renewcommand{\arraystretch}{1.5} % Increase the vertical spacing by 50%
  \caption{\textcolor{black}{Comparison with other methods.}}
  \label{tab:loft_attacks}
  \begin{tabular}{|l|p{1.5cm}|p{1.5cm}|c|p{1.25cm}|}
    \hline
    Method & Mean utilization & Total overflows & CPU & Memory (MB) \\
    \hline
    FloRa & 62.23\% & 1 & 1.86\% & 217 \\
    \hline
    RIT  & 83.98\% & 49 & 0.50\% & 19 \\
    \hline
     LOFTGUARD  & 89.58\% & 17 & 2.11\% & 178 \\
    \hline
     Sphinix  & 84.85\% & 36 & 1.14\% & 49 \\
    \hline

  \end{tabular}
\end{table}
To evaluate the performance of FloRa after deployment, an arrangement with different flow table sizes is made. According to our arrangement, a flow table with a capacity of 1500 entries would have nearly 1500 legitimate flow rules, a 2000 entry capacity flow table would have about 2000 normal flow rules, and about 2500 normal flow rules would be there in the flow table with capacity of 2500 entries and a flow table with 3000 entries capacity would have 3000 normal flow rules. When FloRa is not in use, the LOFT takes place and overwhelms the flow table between $80^{th}$ to $100^{th}$ second in different configurations, as shown in Fig \ref{WF}.
    In the flow table space without Flora, Fig. \ref{WF} shows the count of normal and attack flow rules. Over 80\%\ of the space in the flow tables is occupied by regular flow rules in the first 60 seconds. This shows that, even in a normal network setting, a significant amount of data processing is placed on flow tables. The flow table with a capacity of 1500 entries and 2000 entries continuously overflows beyond the $70^{th}$ second, and the flow table with a capacity of 2500 entries and 3000 entries overflows repeatedly after $80^{th}$ second once the LOFT attack starts. In the $100^{th}$ second, attack flow completely takes over the flow table, consuming the sizable chunk of the area reserved for legitimate flow rules.
    \begin{figure}
    \centering
    \subfigure[ Flow Table size 1500\label{wif1}]{\includegraphics[width=0.45\textwidth]{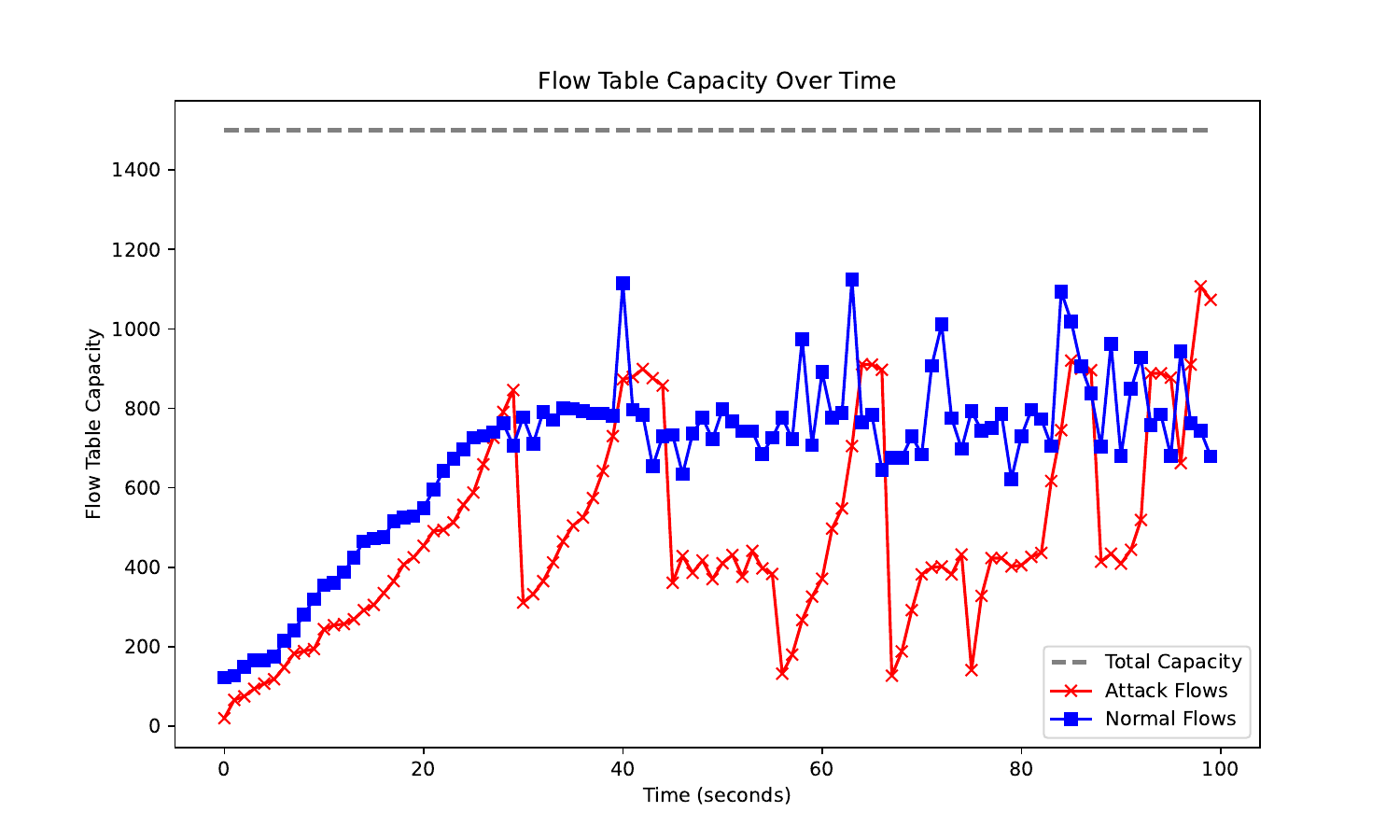}}
    
    \subfigure[Flow Table size 2000\label{wif2}]{\includegraphics[width=0.45\textwidth]{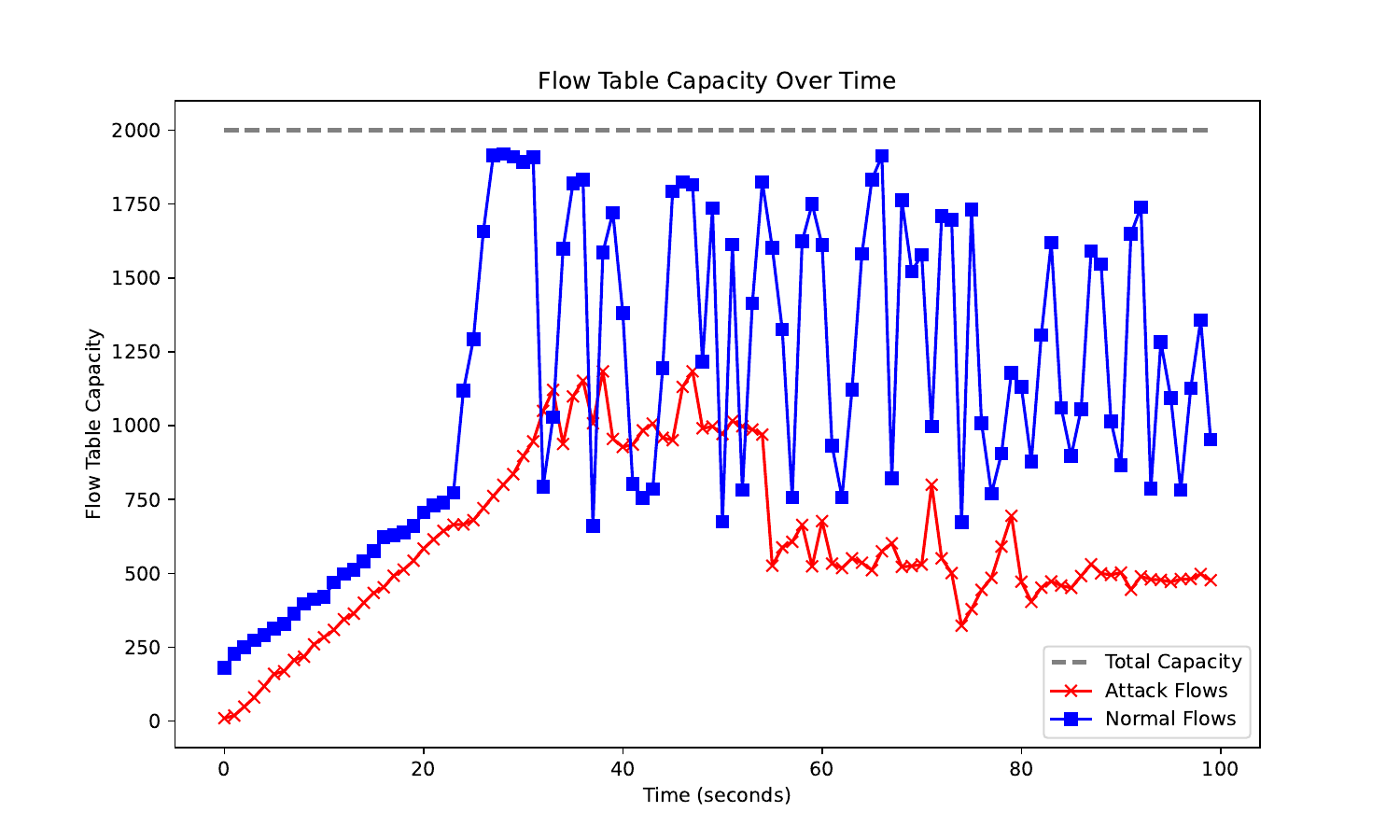}}
            
    \subfigure[Flow Table size 2500\label{wif3}]{\includegraphics[width=0.45\textwidth]{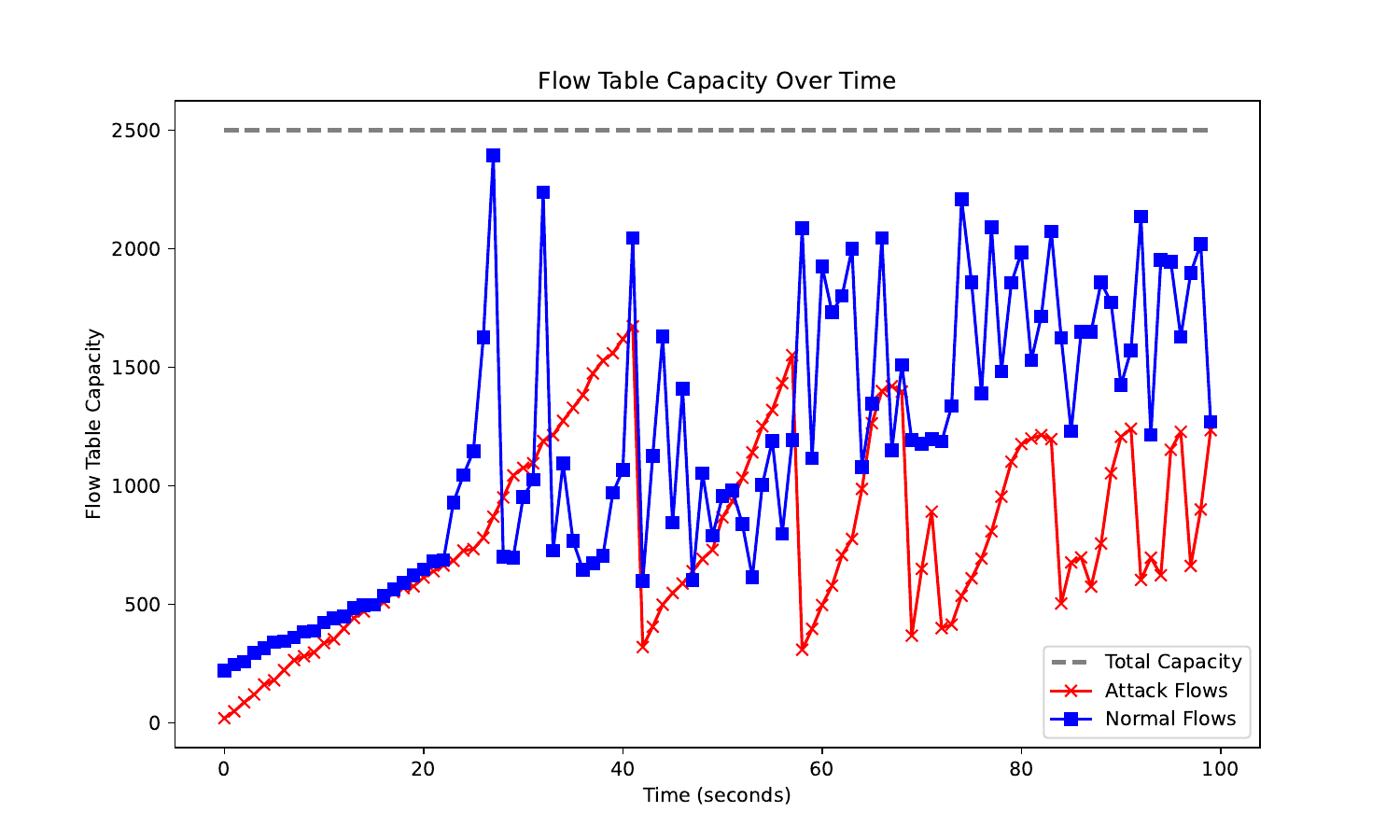}}
    \subfigure[Flow Table size 3000\label{wif4}]{\includegraphics[width=0.45\textwidth]{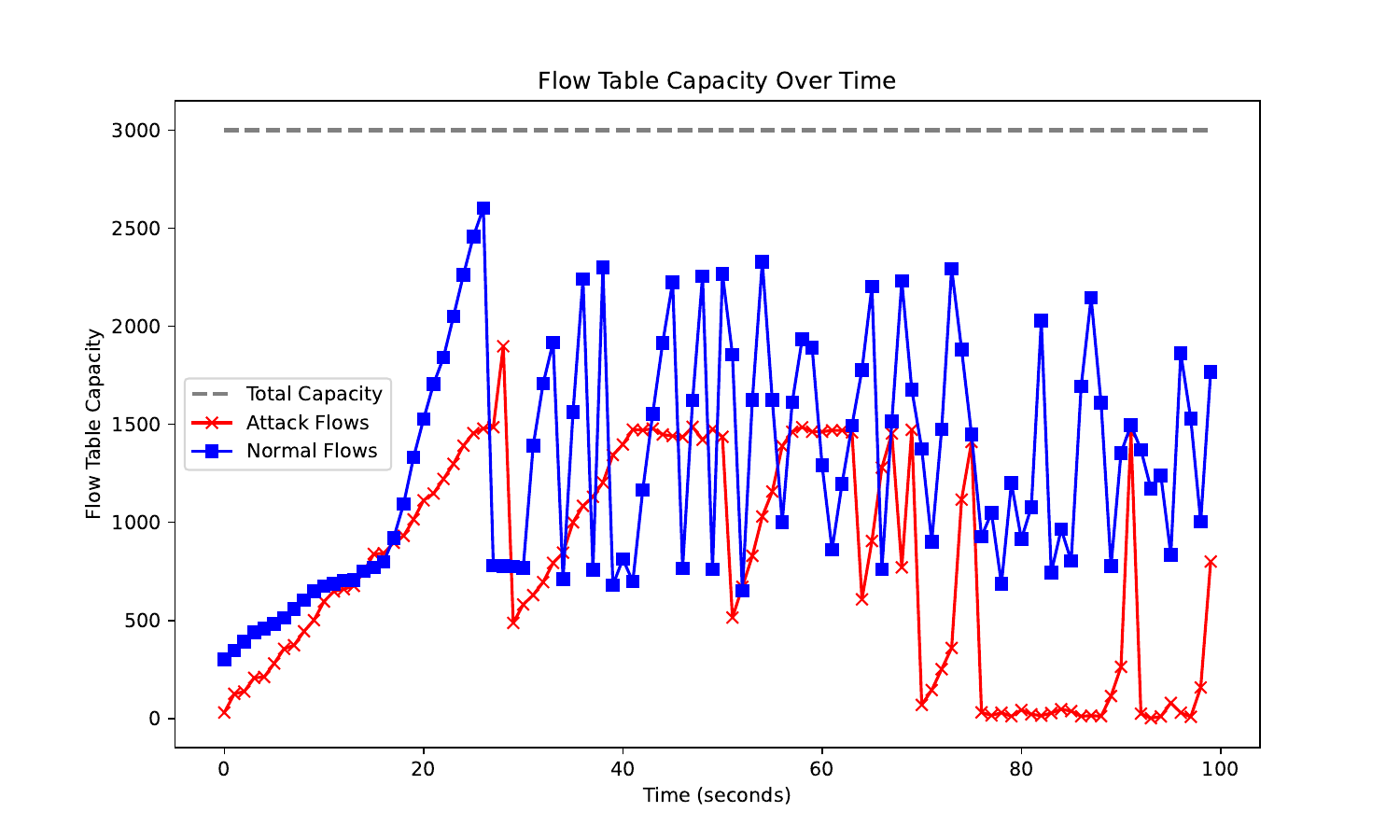}}
    
    \caption{Normal and attack flow rules with FloRa}
    \label{WIF}
    \end{figure}

Fig. \ref{WIF} presents the count of normal and attack flow rules within the flow table space in different arrangements when FloRa is employed. Here, we observe that attack flow rules consistently occupy less than 35\%\ of the flow table space during initial stages, allowing for genuine transmission and handling of most normal flows. In the later stages, FloRa effectively detects the attack flows and sends them to the SDN controller that blocks the attack source immediately, facilitating the flow table regain its regular state by removing attack flows. This shows that FloRa is capable of accurately identifying the majority of attack flows and help remove these rules to allow legitimate rules to be installed and handled properly, also identifies and blacklists the attack source.

\subsection{Analysis of implementation overhead of FloRa}
The performance of FloRa is evaluated based on overhead imposed by the detection logic in terms of CPU and memory usage. In addition, the classification rate is also computed.

\subsubsection{Classification Rate}
\begin{figure}[h]
    \centering
    \subfigure[Classification Rate \label{classification}]{\includegraphics[width=.5\textwidth]{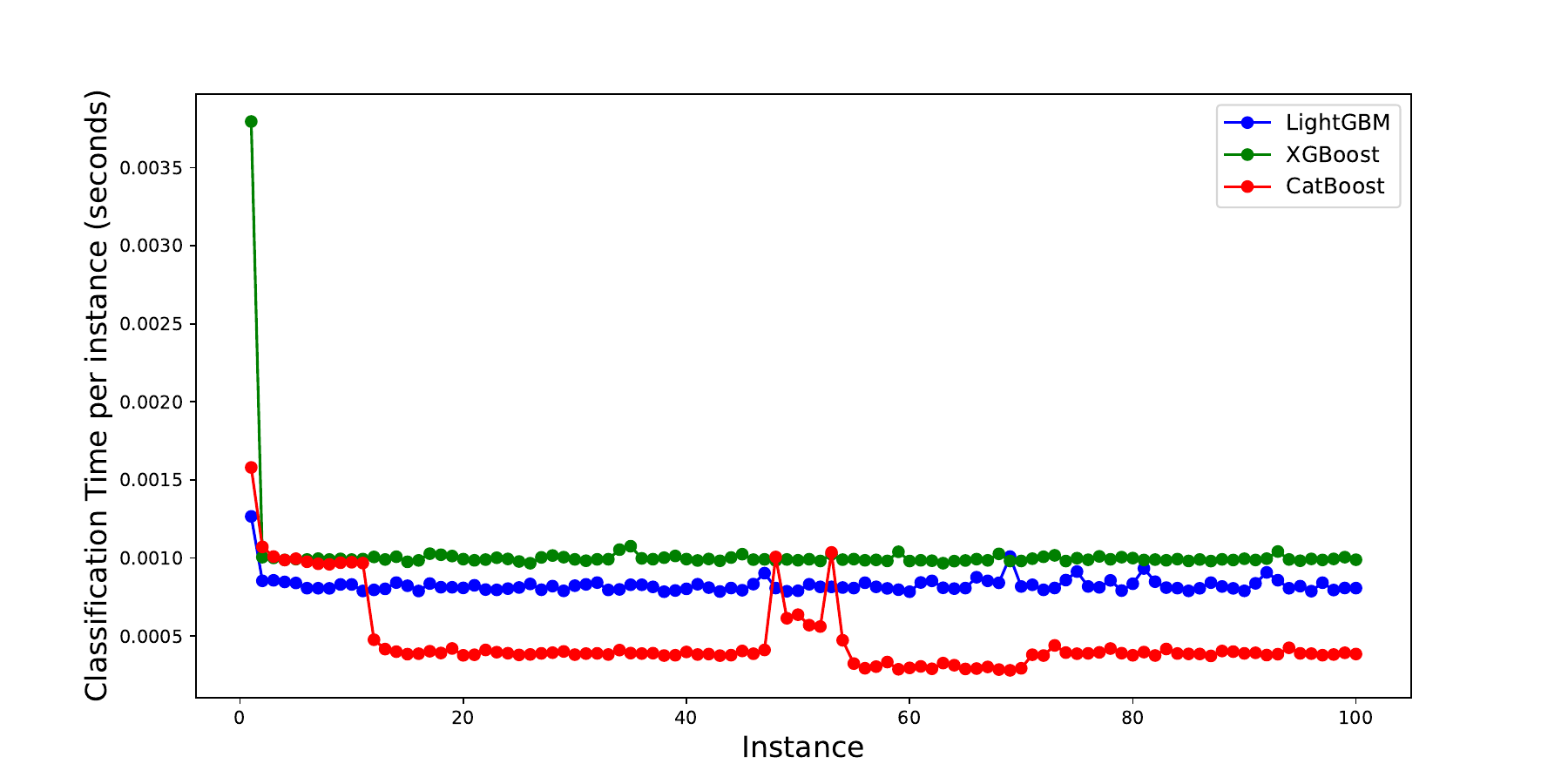}}
    \subfigure[CPU Usage\label{cpu}]{ \includegraphics[width=.5\textwidth]{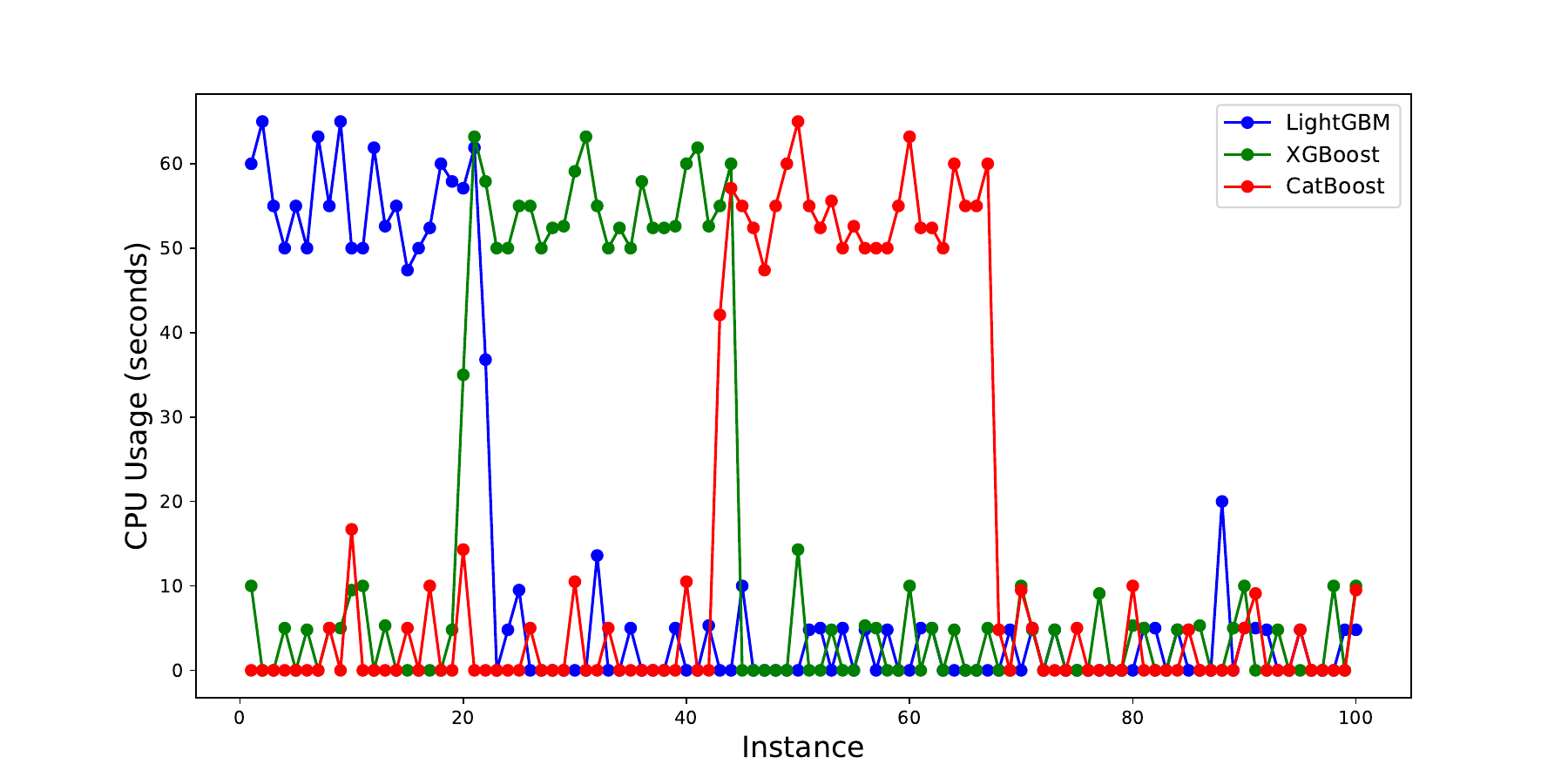}}
    \subfigure[Memory Usage \label{ram}]{\includegraphics[width=.5\textwidth]{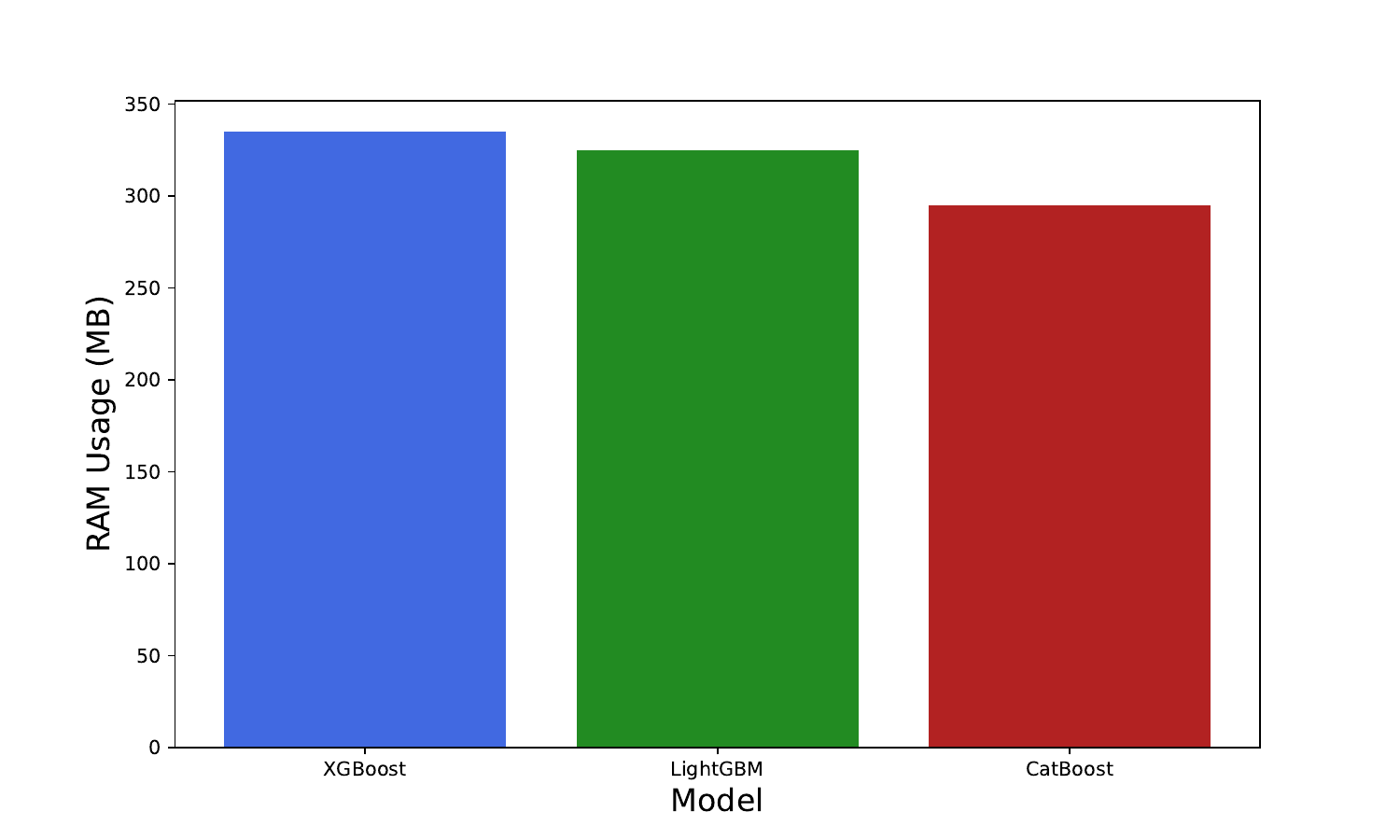}}
   \caption{Comparison of classification rate, CPU usage and Memory usage}
\end{figure}

Fig. \ref{classification} shows the comparative analysis of classification time taken by different classifiers for the classification of each instance in the developed dataset, and it is visible that Cat Boost attains the highest classification rate among all three classifiers. Cat Boost includes options for L1 and L2 regularization, which can help maintain a balance between model complexity and accuracy, maintaining high classification accuracy on unseen data. Cat Boost uses a highly optimized algorithm for constructing DT. It employs techniques like ordered boosting, improving the learning process's overall efficiency.

\subsubsection{CPU overhead}
 The comparison of FloRa with some existing methods is given in Fig. \ref{cpu}. It can be observed that Cat Boost causes the least overhead in comparison to XGBoost and LightGBM. Cat Boost handles categorical features effectively. It is user-friendly and has sensible default hyperparameters; it often performs well in predictive accuracy and is robust against overfitting. It has built-in support for handling imbalanced datasets.

\subsubsection{Memory Overhead}
The existing methods, such as XGBoost, are memory inefficient and consume comparatively high memory, which causes unnecessary overhead for the system and increases the cost. Cat Boost decreases the memory requirement of FloRa in comparison with the existing methods, as indicated in Fig. \ref{ram}. 

\section{Conclusion}\label{cc}
A responsive defense mechanism named FloRa is proposed to secure SDN against LOFT attacks by continuously monitoring flow table rule numbers and real-time prediction of flow table overflow. By analyzing the fundamentals and effects of LOFT attacks, key characteristics that might depict the condition of the flow table have been established. The method successfully identifies LOFT attacks by using feature selection approaches. An attack detection model is trained to categorize flow entries and dynamically determine the fraction of suspicious entries, achieving real-time detection of LOFT attacks. FloRa is evaluated through a set of tests carried out in the Mininet. The prediction performance of FloRa in terms of flow entry counts, attack detection effectiveness, and the ability to spot malicious rules are studied. Additionally, FloRa reduces the classification overhead, CPU, and memory usage.  However, FloRa still suffers from some limitations. It was observed during the offline training that the training time of FloRa was high. To make it feasible to train dynamically, making it capable of detecting attacks with different features, the training time needs to be reduced further. In the future, training time will be considered a critical agenda to make the method capable of detecting other malicious and improved attacks. Moreover, the effectiveness, vulnerability, and limitation of FloRa in real-world deployments will be further evaluated. 

\section*{Acknowledgement}
This work was supported by the Kempe fellowship via project no. SMK21-0061, Sweden. Additional support was provided by the Wallenberg AI, Autonomous Systems and Software Program (WASP) funded by Knut and Alice Wallenberg Foundation and the European Commission through the Horizon Europe project SovereignEdge.COGNIT (grant no. 101092711).

\section*{Note to Readers}
This is a preprint of an article published in  IEEE Transactions on Network and Service Management. The final version of this paper is available at: \href{https://doi.org/10.1109/TNSM.2024.3446178}{https://doi.org/10.1109/TNSM.2024.3446178}.


%\begin{thebibliography}{1}


\bibliographystyle{IEEEtran}

\vfill


\end{document}